\documentclass[aps,prd,showpacs,showkeys,amsmath,amssymb]{revtex4}
\usepackage{amsfonts}
\usepackage{graphicx}
\usepackage{amsmath}
\usepackage{mathrsfs}
\usepackage{dcolumn}
\usepackage{bm}
\usepackage{array}
\usepackage{caption2}
\allowdisplaybreaks
\begin{document}
\title{Strong and Electromagnetic Decays of The $D$-wave Heavy Mesons}
\author{Peng-Zhi Huang}
\email{pzhuang@pku.edu.cn}
\author{Liang Zhang}
\email{liangzhang@pku.edu.cn} \affiliation{Department of Physics
and State Key Laboratory of Nuclear Physics and Technology\\
Peking University, Beijing 100871, China  }
\author{Shi-Lin Zhu}
\email{zhusl@pku.edu.cn} \affiliation{Department of Physics
and State Key Laboratory of Nuclear Physics and Technology\\
and Center of High Energy Physics, Peking University, Beijing
100871, China }

\begin{abstract}

We calculate the $\pi$, $\rho$, $\omega$, and $\gamma$ coupling
constants between the heavy meson doublets $(1^-,2^-)$ and
$(0^-,1^-)$/$(0^+,1^+)$ within the framework of the light-cone QCD
sum rule at the leading order of heavy quark effective theory.
Most of the sum rules are stable with the variations of the Borel
parameter and the continuum threshold. Then we calculate the
strong and electromagnetic decay widths of the $(1^-,2^-)$
$D$-wave heavy mesons. Their total widths are around several tens
of MeV, which is helpful in the future experimental search.

\end{abstract}
\maketitle
\pagenumbering{arabic}

%%%%%%%%%%%%%%%%%%%%%%%%%%%%%%%%%%%%%%%%%%%%%%
\section{Introduction}\label{introduction}
%%%%%%%%%%%%%%%%%%%%%%%%%%%%%%%%%%%%%%%%%%%%%

Heavy quark effective theory (HQET) \cite{hqet} is a framework
which is widely used to study the spectra and transition
amplitudes of heavy hadrons containing one heavy quark. In HQET,
the expansion is performed in terms of $1/m_Q$, where $m_Q$ is the
mass of the heavy quark involved. At the leading order of $1/m_Q$,
the HQET Lagrangian respects the heavy quark flavor-spin symmetry,
therefore heavy hadrons form a series of degenerate doublets. The
two members in a doublet share the same quantum number $j_l$, the
angular momentum of the light components. The $j_l={1 \over 2}$
$S$-wave doublet $(0^-,1^-)$ is conventionally denoted as $H$ and
the $j_l={1\over 2}/{3 \over 2}$ $P$-wave doublets
$(0^+,1^+)/(1^+,2^+)$ are conventionally denoted as $S/T$. We
denote the $j_l={3\over 2}/{5 \over 2}$ $D$-wave doubtlets
$(1^-,2^-)/(2^-,3^-)$ as $M/N$.

Shifman-Vainshtein-Zakharov (SVZ) sum rules \cite{svz} is a
nonperturbative approach used to determine hadronic parameters
such as the hadron mass. The vacuum expectation value of the $T$
product of two interpolating currents is considered in this
approach. After performing the operator product expansion (OPE),
one obtains sum rules which relate the hadronic parameters to
expressions containing vacuum condensates parameterizing the QCD
nonperturbative effect. In the late 1980s, light-cone QCD sum
rules (LCQSR) \cite{light-cone} was developed to calculate various
hadronic transition form factors. Now the OPE of the $T$ product
of two interpolating currents sandwiched between the vacuum and an
hadronic state is performed near the light-cone rather than at a
small distance as in the conventional SVZ sum rules.

The $\rho$ coupling constants $g_{B^*B\rho}$ and $g_{D^*D\rho}$ were calculated
with LCQSR in full QCD in Ref. \cite{rhodecayaliev}. The couplings
$g_{H*H*\rho}$, $f_{H*H*\rho}$, $g_{HH\rho}$, and $f_{H*H\rho}$
were calculated in full QCD in Ref. \cite{rhodecayli}. Their values in the
limit $m_Q\rightarrow \infty$ are also discussed in this paper.
The $\rho$ coupling constants between the three doublets $H$, $S$, $T$ and within the two
doublets $H$, $S$ are systematically studied with LCQSR at the
leading order of HQET in Ref. \cite{rhodecayzhu}.
The $\pi$ coupling constants between the $S$-wave and $P$-wave heavy mesons have been
studied using QCD sum rules or/and LCQSR in Ref. \cite{pidecaycolangelo}.
The $\pi$ coupling constants between $M$/$N$ and $H$/$S$/$T$
are calculated with LCQSR at the leading order of HQET in Ref. \cite{pidecaydaizhu,pidecaywei}.
The radiative decay between $H$, $S$, and $T$ are studied using the light-cone QCD sum rule
at the leading order of HQET in Ref. \cite{radiativedecayzhu}.
In Ref. \cite{radiativedecaycolangelo}, the radiative decays of $D^*_{sJ}(2317)$ and $D_{sJ}(2460)$
are studied using LCQSR approach.

In this work, we use LCQSR to calculate the $\pi$, $\rho$, $\omega$, and $\gamma$ coupling
constants between the doublets $M$ and $H$/$S$. Because of the covariant derivative in the
interpolating currents of the $M$ doublet, the contribution from the
3-particle light-cone distribution amplitudes of $\pi$, $\rho$, $\omega$, and $\gamma$
have to be included. We work in HQET to differentiate the
two states with the same $J^P$ value and yet quite different decay
widths. The interpolating currents
$J^{\alpha_1\cdots\alpha_j}_{j,P,j_l}$ adopted in our work have
been properly constructed in Ref. \cite{huang}. They satisfy
\begin{eqnarray}
\langle 0|J^{\alpha_1\cdots\alpha_j}_{j,P,j_l}(0)|j',P',j'_l\rangle
&=&f_{Pj_l}\delta_{jj'}\delta_{PP'}\delta_{j_lj'_l}\eta^{\alpha_1\cdots\alpha_j}\label{eq:OverlapAmp}\,,\\
i\langle 0|T\{J^{\beta_1\cdots\beta_{j'}}_{j,P,j_l}(x)J^{\dag\alpha_1\cdots\alpha_j}_{j',P',j'_l}(0)\}|0\rangle
&=&\delta_{jj'}\delta_{PP'}\delta_{j_lj'_l}(-1)^j \mathcal {S}g_t^{\alpha_1\beta_1}\cdots g_t^{\alpha_j\beta_j}
\int dt\delta(x-vt)\Pi_{P,j_l}(x)\,,
\end{eqnarray}
in the limit $m_Q\rightarrow \infty$.
Here $\eta^{\alpha_1\cdots\alpha_j}$ is the polarization tensor for the spin $j$ state,
$v$ is the velocity of the heavy quark, $g_t^{\alpha\beta}=g^{\alpha\beta}-v^\alpha v^\beta$,
$\mathcal {S}$ denotes symmetrizing the indices and subtracting the trace terms separately in the sets
$(\alpha_1\cdots\alpha_j)$ and $(\beta_1\cdots\beta_j)$.

%%%%%%%%%%%%%%%%%%%%%%%%%%%%%%%%%%%%%%%%%%%%%%%%%%%%%%%%%%%%%%%
\section{Sum Rules for the $\pi$ coupling constants}\label{picoupling}
%%%%%%%%%%%%%%%%%%%%%%%%%%%%%%%%%%%%%%%%%%%%%%%%%%%%%%%%%%%%%%%

We shall perform the calculation at the leading order of HQET.
According to Ref. \cite{huang}, the interpolating currents for
the doublets $(0^-,1^-)$, $(0^+,1^+)$, and $(1^-,2^-)$ read as
\begin{eqnarray}
J^\dag_{0,-,\frac{1}{2}}&=&\sqrt{\frac{1}{2}}\bar{h}_v\gamma_5 q\,,\nonumber\\
J^{\dag\alpha}_{1,-,\frac{1}{2}}&=&\sqrt{\frac{1}{2}}\bar{h}_v\gamma_t^\alpha q\,,\nonumber\\
J^\dag_{0,+,\frac{1}{2}}&=&\sqrt{\frac{1}{2}}\bar{h}_v q\,,\nonumber\\
J^{\dag\alpha}_{1,+,\frac{1}{2}}&=&\sqrt{\frac{1}{2}}\bar{h}_v\gamma_5\gamma_t^\alpha q\,,\nonumber\\
J^{\dag\alpha}_{1,+,\frac{3}{2}}
&=&\sqrt{\frac{3}{4}}\bar{h}_v\gamma_5(-i)\left\{\mathcal {D}_t^\alpha-\frac{1}{3}\gamma_t^\alpha\hat{\mathcal {D}}_t\right\}q\,,\nonumber\\
J^{\dag\alpha_1\alpha_2}_{2,+,\frac{3}{2}}
&=&\sqrt{\frac{1}{8}}\bar{h}_v(-i)\left\{\gamma_t^{\alpha_1}\mathcal {D}_t^{\alpha_2}
+\gamma_t^{\alpha_2}\mathcal {D}_t^{\alpha_1}-\frac{2}{3}g_t^{\alpha_1\alpha_2}\hat{\mathcal {D}}_t\right\}q\,,\nonumber\\
J^{\dag\alpha}_{1,-,{3\over 2}}
&=&\sqrt{\frac{3}{4}}\bar{h}_v(-i)\left\{\mathcal {D}_t^{\alpha}-\frac{1}{3}\gamma_t^{\alpha}\hat{\mathcal {D}}_t\right\}q\,,\nonumber\\
J^{\dag\alpha_1\alpha_2}_{2,-,\frac{3}{2}}
&=&\sqrt{\frac{1}{8}}\bar{h}_v(-i)\gamma_5\left\{\gamma_t^{\alpha_1}\mathcal {D}_t^{\alpha_2}
+\gamma_t^{\alpha_2}\mathcal {D}_t^{\alpha_1}-\frac{2}{3}g_t^{\alpha_1\alpha_2}\hat{\mathcal {D}}_t\right\}q\,,
\end{eqnarray}
where $h_v$ is the heavy quark field in HQET, $\gamma_t^\mu\equiv \gamma^\mu-\hat{v}v^\mu$,
$\mathcal {D}_t^\mu\equiv \mathcal {D}^\mu-(\mathcal {D}\cdot v)v^\mu$,
$g_t^{\mu\nu}\equiv g^{\mu\nu}-v^\mu v^\nu$, and $v^\mu$ is the velocity of the heavy quark.

We consider the $\pi$ decay of $M_2$ to $H_1$ to illustrate our
calculation. Here the subscript of $M$($H$) indicates the spin of
the meson involved. Owing to the conservation of the angular
momentum of the light components in the limit $m_Q\rightarrow \infty$,
there is only one independent $\pi$ coupling constant between
doublets $M$ and $H$. We denote it as $g^{p1}_{MH\pi}$ where $p$ and
the number following it indicate the orbital and total angular
momentum $(l,j_h)$ of the final $\pi$ meson respectively.
$g^{p1}_{MH\pi}$ can be defined in terms of the decay amplitude
$\mathcal {M}(M_2\rightarrow H_1+\pi)$ as
\begin{eqnarray}
\mathcal {M}(M_2\rightarrow H_1+\pi)
&=&I\eta_{\alpha_1\alpha_2}[\epsilon^{*\alpha_1}_tq^{\alpha_2}_t
-\frac{1}{3}g_t^{\alpha_1\alpha_2}(\epsilon^*\cdot q_t)]g^{p1}_{M_2H_1\pi}\,,
\end{eqnarray}
where $\eta$ and $\epsilon^*$ denote the polarization tensors of
the initial and final heavy mesons respectively, $q$ is the
momentum of the $\pi$ meson. The transversal tensor
are defined as $\epsilon^\mu_t\equiv \epsilon^\mu-(\epsilon\cdot
v)v^\mu$, $q^\mu_t\equiv q^\mu-(q\cdot v)v^\mu$, and
$g_t^{\mu\nu}\equiv g^{\mu\nu}-v^\mu v^\nu$. $I=1,1/\sqrt{2}$ for
the charged and neutral $\pi$ meson, respectively.

To obtain the sum rules for the coupling constants
$g^{p1}_{M_2H_1\pi}$, we consider the correlation function
\begin{eqnarray} \label{eq:CorHadpi}
\int d^4 xe^{-ik\cdot x}\langle \pi(q)|T\{J^\beta_{1,-,\frac{1}{2}}(0)J^{\dag\alpha_1\alpha_2}_{2,-,\frac{3}{2}}(x)\}|0\rangle
&=&I\left[\frac{1}{2}(g_t^{\alpha_1\beta}q_t^{\alpha_2}+g_t^{\alpha_2\beta}q_t^{\alpha_1})
-\frac{1}{3}g_t^{\alpha_1\alpha_2}q_t^\beta\right]G^{p1}_{M_2H_1\pi}(\omega,\omega')\,,
\end{eqnarray}
where $\omega\equiv 2v\cdot k$, $\omega'\equiv 2v\cdot(k-q)$.
At the leading order of HQET, the heavy quark propagator reads as
\begin{eqnarray}
\langle 0|T\{h_v(0)\bar{h}_v(x)\}|0\rangle=\frac{1+\hat{v}}{2}\int dt\delta^4(-x-vt)\,.
\end{eqnarray}
The correlation function can now be expressed as
\begin{eqnarray}
-\frac{i}{4}
\int dx e^{-ik\cdot x}\int_0^\infty dt \delta(-x-vt)
\text{Tr}\biggl\lbrace\gamma_t^\beta\frac{1+\hat{v}}{2}\gamma_5
\left[\gamma_t^{\alpha_1}\mathcal {D}_t^{\alpha_2}+\gamma_t^{\alpha_2}\mathcal {D}_t^{\alpha_1}
-\frac{2}{3}g_t^{\alpha_1\alpha_2}\hat{\mathcal {D}}_t\right]
\langle\pi(q)|q(x)\bar{q}(0)|0\rangle\biggl\rbrace\,.
\end{eqnarray}
It can be further calculated using the light-cone wave functions
of the $\pi$ meson. To our approximation,
we need the 2- and 3-particle light-cone wave functions. Their definitions are
collected in Appendix~\ref{appendixLCDA}.

At the hadron level, $G^{p1}_{M_2H_1\pi}(\omega,\omega')$ in (\ref{eq:CorHadpi}) has the following pole terms
\begin{eqnarray}\label{eq:PoleTermpi}
G^{p1}_{M_2H_1\pi}(\omega,\omega')
=\frac{f_{-,1/2}f_{-,3/2}g^{p1}_{M_2H_1\pi}}{(2\bar{\Lambda}_{-,1/2}-\omega')(2\bar{\Lambda}_{-,3/2}-\omega)}
+\frac{c}{2\bar{\Lambda}_{-,1/2}-\omega'}+\frac{c'}{2\bar{\Lambda}_{-,3/2}-\omega}\,,
\end{eqnarray}
where $\bar{\Lambda}_{-,1/2}\equiv m_H-m_Q$,
$\bar{\Lambda}_{-,3/2}\equiv m_M-m_Q$, $f_{-,1/2}$, etc. are the
overlap amplitudes of their interpolating currents with the heavy
mesons.

$G^{p1}_{M_2H_1\rho}(\omega,\omega')$ can now be expressed by the
$\pi$ meson light-cone wave functions. After the Wick rotation
and the double Borel transformation with $\omega$ and $\omega'$,
the single-pole terms in (\ref{eq:PoleTermpi}) are eliminated.
We arrive at
\begin{eqnarray}\label{eq:SumRulepi1}
&&g^{p1}_{M_2H_1\pi}f_{-,\frac{1}{2}}f_{-,\frac{3}{2}}
e^{-\frac{\bar{\Lambda }_{-,3/2}+\bar{\Lambda }_{-,1/2}}{T}}\nonumber\\
&&=
-\frac{1}{48}f_\pi\biggl\lbrace
-12\left[\phi_\pi'(\bar{u}_0)
-(u\phi_\pi)'(\bar{u}_0)\right]T^2f_1(\frac{\omega_c}{T})\nonumber\\
&&\mathrel{\phantom{=}}
-\frac{4m_\pi^2}{m_u+m_d}\left[6\mathcal{T}^{[1,0]}(u_0)
+6\phi_p(\bar{u}_0)
-6(u\phi_p)(\bar{u}_0)
+\phi_\sigma(\bar{u}_0)\right]Tf_0(\frac{\omega_c}{T})\nonumber\\
&&\mathrel{\phantom{=}}
+3m_\pi^2\left[\mathbb{A}'(\bar{u}_0)
-(u\mathbb{A})'(\bar{u}_0)
+8\mathbb{B}^{[2]}(\bar{u}_0)
-8\mathbb{B}^{[1]}(\bar{u}_0)
+8(u\mathbb{B})^{[1]}(\bar{u}_0)
-16\mathcal{V}^{[0,0]}_\perp(u_0)-16\mathcal{A}^{[0,0]}_\parallel(u_0)\right]\biggl\rbrace\,,
\end{eqnarray}
where $f_n(x)=1-e^{-x}\sum_{k=0}^{n}x^k/k!$ is the
continuum subtraction factor, and $\omega_c$ is the continuum
threshold, $u_0=T_1/(T_1+T_2)$, $T=T_1T_2/(T_1+T_2)$,
and $\bar{u}_0 = 1-u_0$. $T_1$ and $T_2$ are the two Borel
parameters. We have employed the Borel transformation $\widetilde{\mathcal
{B}}_\omega^Te^{\alpha\omega}=\delta(\alpha-1/T)$ to
obtain (\ref{eq:SumRulepi1}). In the above expressions, we have used the
functions $\mathcal {F}^{[a]}(\bar{u}_0)$ and $\mathcal
{F}^{[a,b]}(u_0)$ which are defined in Appendix~\ref{appendixF}.

The $\pi$ coupling constant between doublets $M$ and $S/T/M$ can be defined similarly:
\begin{eqnarray}\label{piamplitude}
\mathcal {M}(M_2\rightarrow S_1+\pi)
&=&Ii\eta_{\alpha_1\alpha_2}\epsilon^*_\beta
\varepsilon^{\beta\alpha_1qv}q_t^{\alpha_2}g^{d2}_{M_2S_1\pi}\,,\nonumber
\\
\mathcal {M}(M_1\rightarrow T_1+\pi)
&=&I(\eta\cdot\epsilon^*_t)g^{s0}_{M_1T_1\pi}
+I\left[(\eta\cdot q_t)(\epsilon^*\cdot q_t)-\frac{q_t^2}{3}(\eta\cdot\epsilon^*_t)\right]g^{d2}_{M_1T_1\pi}\,,\nonumber
\\
\mathcal {M}(M_2\rightarrow M_1+\pi)
&=&2I\eta_{\alpha_1\alpha_2}\left[\epsilon^{*\alpha_1}_tq^{\alpha_2}_t-\frac{1}{3}g_t^{\alpha_1\alpha_2}(\epsilon^*\cdot q_t)\right]g^{p1}_{M_2M_1\pi}\nonumber\\
&&+I\eta_{\alpha_1\alpha_2}\left\{q^{\alpha_1}_tq^{\alpha_2}_t(\epsilon^*\cdot q_t)
-\frac{q_t^2}{5}\left[2\epsilon^{*\alpha_1}_tq^{\alpha_2}_t+g_t^{\alpha_1\alpha_2}(\epsilon^*\cdot q_t)\right]\right\}g^{f3}_{M_2M_1\pi}\,.
\end{eqnarray}

Here the vector notations in the Levi-Civita tensor come from an index contraction
between the Levi-Civita tensor and the vectors,
for example, $\varepsilon^{\alpha\beta qv}\equiv\varepsilon^{\alpha\beta\gamma\delta}
q_\gamma v_\delta$. The Levi-Civita tensor is defined as $\varepsilon^{0123}=1$.
The sum rules for the coupling constants in Eq. (\ref{piamplitude}) are
\begin{eqnarray}\label{eq:SumRulepi2}
&&g^{d2}_{M_2S_1\pi}f_{+,\frac{1}{2}}f_{-,\frac{3}{2}}
e^{-\frac{\bar{\Lambda }_{-,3/2}+\bar{\Lambda }_{+,1/2}}{T}}\nonumber\\
&&=
-\frac{1}{24}f_\pi\biggl\lbrace
12\left[\phi_\pi(\bar{u}_0)
-(u\phi_\pi)(\bar{u}_0)\right]Tf_0(\frac{\omega_c}{T})
+\frac{4m_\pi^2}{m_u+m_d}\left[6\mathcal{T}^{[0,0]}(u_0)
-\phi_\sigma(\bar{u}_0)
+(u\phi_\sigma)(\bar{u}_0)\right]\nonumber\\
&&\mathrel{\phantom{=}}
-3m_\pi^2\left[\mathbb{A}(\bar{u}_0)
-(u\mathbb{A})(\bar{u}_0)
+16\mathcal{A}^{[-1,0]}_\parallel(u_0)
+16\mathcal{A}^{[-1,0]}_\perp(u_0)\right]\frac{1}{T}\biggl\rbrace\,,\nonumber
\\
&&g^{s0}_{M_1T_1\pi}f_{+,\frac{3}{2}}f_{-,\frac{3}{2}}e^{-\frac{\bar{\Lambda }_{-,3/2}+\bar{\Lambda }_{+,3/2}}{T}}\nonumber
\\
&&=\frac{f_\pi}{48}(u(1-u)\phi_\pi)^{(3)}(\bar{u}_0)T^4f_3(\frac{\omega_c}{T})+\frac{f_\pi m_\pi^2}{24m_{ud}}
\left[\frac{m_\pi^2-m_{ud}^2}{6m_\pi^2}\phi_\sigma^{(2)}(\bar{u}_0)
+(u(1-u)\phi_P)^{(2)}(\bar{u}_0)+(1-\alpha_2\mathcal{T})^{[3,0]}(u_0)\right.\nonumber
\\
&&\mathrel{\phantom{=}}\left.
-(\alpha_3\mathcal{T})^{[3,1]}(u_0)\right]T^3f_2(\frac{\omega_c}{T})-\frac{f_\pi m_\pi^2}{12}
\left[\frac{3}{8}\mathbb{A}'(\bar{u}_0)+\frac{1}{16}(u(1-u)\mathbb{A})^{(3)}(\bar{u}_0)
+\mathbb{B}^{[1]}(\bar{u}_0)+(u\mathbb{B}-\frac{1}{2}\mathbb{B})(\bar{u}_0)\right.\nonumber
\\
&&\mathrel{\phantom{=}}
-\frac{1}{2}(u(1-u)\mathbb{B})'(\bar{u}_0)+(u(1-u)\phi_\pi)'(\bar{u}_0)
+(2\mathcal{A}_\parallel+2\mathcal{A}_\perp+\mathcal{V}_\parallel+\mathcal{V}_\perp)^{[1,0]}(u_0)
+((1-\alpha_2)(\mathcal{A}_\parallel+\mathcal{V}_\perp))^{[2,0]}(u_0)\nonumber
\\
&&\mathrel{\phantom{=}}\left.
-(\alpha_3(\mathcal{A}_\parallel+\mathcal{V}_\perp))^{[2,1]}(u_0)\right]T^2f_1(\frac{\omega_c}{T})
-\frac{f_\pi m_\pi^4}{6m_{ud}}\left[(u(1-u)\phi_P)(\bar{u}_0)
+\frac{m_\pi^2-m_{ud}^2}{6m_\pi^2}\phi_\sigma(\bar{u}_0)
+((1-\alpha_2)\mathcal{T})^{[1,0]}(u_0)\right.\nonumber
\\
&&\mathrel{\phantom{=}}\left.
-(\alpha_3\mathcal{T})^{[1,1]}(u_0)\right]Tf_0(\frac{\omega_c}{T})
+\frac{f_\pi m_\pi^4}{3}\left[\frac{1}{16}(u(1-u)\mathbb{A})'(\bar{u}_0)
+(u\mathbb{B}-\frac{1}{2}\mathbb{B})^{[2]}(\bar{u}_0)
-\frac{1}{2}(u(1-u)\mathbb{B})^{[1]}(\bar{u}_0)\right.\nonumber
\\
&&\mathrel{\phantom{=}}\left.
+\mathbb{B}^{[3]}(\bar{u}_0)
+(\mathcal{A}_\parallel+\mathcal{A}_\perp-\mathcal{V}_\parallel-\mathcal{V}_\perp)^{[-1,0]}(u_0)
-((1-\alpha_2)(\mathcal{A}_\parallel+\mathcal{V}_\perp))^{[0,0]}(u_0)
+(\alpha_3(\mathcal{A}_\parallel+\mathcal{V}_\perp))^{[0,1]}(u_0)\right]\,,\nonumber
\\
&&g^{d2}_{M_1T_1\pi}f_{+,\frac{3}{2}}f_{-,\frac{3}{2}}e^{-\frac{\bar{\Lambda }_{-,3/2}+\bar{\Lambda }_{+,3/2}}{T}}\nonumber
\\
&&=-\frac{f_\pi}{8}(u(1-u)\phi_\pi)'(\bar{u}_0)T^2f_1(\frac{\omega_c}{T})
-\frac{f_\pi m_\pi^2}{4m_{ud}}\left[(u(1-u)\phi_P)(\bar{u}_0)
+\frac{m_\pi^2-m_{ud}^2}{6m_\pi^2}\phi_\sigma(\bar{u}_0)
+((1-\alpha_2)\mathcal{T})^{[1,0]}(u_0)\right.\nonumber
\\
&&\mathrel{\phantom{=}}\left.
-(\alpha_3\mathcal{T})^{[1,1]}(u_0)\right]Tf_0(\frac{\omega_c}{T})
+\frac{f_\pi m_\pi^2}{2}\left[\frac{1}{16}(u(1-u)\mathbb{A})'(\bar{u}_0)
-\frac{1}{2}(u(1-u)\mathbb{B})^{[1]}(\bar{u}_0)
+\mathbb{B}^{[3]}(\bar{u}_0)+((u-\frac{1}{2})\mathbb{B})^{[2]}(\bar{u}_0)\right.\nonumber
\\
&&\mathrel{\phantom{=}}\left.
+(\mathcal{A}_\parallel+\mathcal{A}_\perp+2\mathcal{V}_\parallel+2\mathcal{V}_\perp)^{[-1,0]}(u_0)
-((1-\alpha_2)(\mathcal{A}_\parallel+\mathcal{V}_\perp))^{[0,0]}(u_0)
+(\alpha_3(\mathcal{A}_\parallel+\mathcal{V}_\perp))^{[0,0]}(u_0)\right]\,,\nonumber
\\
&&g^{p1}_{M_2M_1\pi}f_{-,\frac{3}{2}}^2e^{-\frac{2\bar{\Lambda }_{-,3/2}}{T}}\nonumber
\\
&&=\frac{f_\pi m_{ud}}{80\sqrt{6}}(u(1-u)\phi_\pi)^{(2)}(\bar{u}_0)T^3f_2(\frac{\omega_c}{T})
-\frac{f_\pi m_\pi^2}{240\sqrt{6}m_{ud}}
\left[\frac{m_\pi^2-m_{ud}^2}{m_\pi^2}(u(1-u)\phi_\sigma)^{(2)}(\bar{u}_0)
+6((1-\alpha_2)\mathcal{T})^{[2,0]}(u_0)\right.\nonumber
\\
&&\mathrel{\phantom{=}}\left.
-6(\alpha_3\mathcal{T})^{[2,1]}(u_0)\right]T^2f_1(\frac{\omega_c}{T})
+\frac{f_\pi m_\pi^2}{20\sqrt{6}}\left[-\frac{5}{8}\mathbb{A}(\bar{u}_0)
-\frac{1}{16}(u(1-u)\mathbb{A})^{(2)}(\bar{u}_0)-(u(1-u)\phi_\pi)(\bar{u}_0)\right.\nonumber
\\
&&\mathrel{\phantom{=}}\left.
-4(\mathcal{A}_\parallel+\mathcal{A}_\perp)^{[0,0]}(u_0)
+((1-\alpha_2)(\mathcal{A}_\parallel+\mathcal{A}_\perp))^{[1,0]}(u_0)
-(\alpha_3(\mathcal{A}_\parallel+\mathcal{A}_\perp))^{[1,1]}(u_0)
-5(\mathcal{V}_\parallel+\mathcal{V}_\perp)^{[0,0]}(u_0)\right]\times\nonumber
\\
&&\mathrel{\phantom{=}}Tf_0(\frac{\omega_c}{T})
+\frac{f_\pi m_\pi^4}{60\sqrt{6}m_{ud}}\left[\frac{m_\pi^2-m_{ud}^2}{m_\pi^2}(u(1-u)\phi_\sigma)(\bar{u}_0)
-6((1-\alpha_2)\mathcal{T})^{[0,0]}(u_0)
+6(\alpha_3\mathcal{T})^{[0,1]}(u_0)\right]
+\frac{f_\pi m_\pi^4}{5\sqrt{6}}\times
\nonumber
\\
&&\mathrel{\phantom{=}}
\left[\frac{1}{16}(u(1-u)\mathbb{A})(\bar{u}_0)
-(\mathcal{A}_\parallel+\mathcal{A}_\perp)^{[-2,0]}(u_0)+((1-\alpha_2)(\mathcal{A}_\parallel+\mathcal{A}_\perp))^{[-1,0]}(u_0)
-(\alpha_3(\mathcal{A}_\parallel+\mathcal{A}_\perp))^{[-1,1]}(u_0)\right]\frac{1}{T}\,,\nonumber
\\
&&g^{f3}_{M_2M_1\pi}f_{-,\frac{3}{2}}^2e^{-\frac{2\bar{\Lambda }_{-,3/2}}{T}}\nonumber
\\
&&=-\frac{\sqrt{6}f_\pi}{4}(u(1-u)\phi_\pi)(\bar{u}_0)Tf_0(\frac{\omega_c}{T})
+\frac{\sqrt{6}f_\pi m_\pi^2}{2m_{ud}}\left[\frac{m_\pi^2-m_{ud}^2}{6m_\pi^2}(u(1-u)\phi_\sigma)(\bar{u}_0)
-((1-\alpha_2)\mathcal{T})^{[0,0]}(u_0)\right.\nonumber
\\
&&\mathrel{\phantom{=}}\left.
+(\alpha_3\mathcal{T})^{[0,1]}(u_0)\right]
+\sqrt{6}f_\pi m_\pi^2\left[\frac{1}{16}(u(1-u)\mathbb{A})(\bar{u}_0)
-(\mathcal{A}_\parallel+\mathcal{A}_\perp)^{[-2,0]}(u_0)
+((1-\alpha_2)(\mathcal{A}_\parallel+\mathcal{A}_\perp))^{[-1,0]}(u_0)\right.\nonumber
\\
&&\mathrel{\phantom{=}}\left.
-(\alpha_3(\mathcal{A}_\parallel+\mathcal{A}_\perp))^{[-1,1]}(u_0)\right]\frac{1}{T}\,,
\end{eqnarray}
with $m_{ud}=m_u+m_d$.
The $\pi$ coupling constants of the other decay channels are defined as
\begin{eqnarray}
\mathcal {M}(M_1\rightarrow H_0+\pi)
&=&I(\eta\cdot q_t)g^{p1}_{M_1H_0\pi}\,,\nonumber
\\
\mathcal {M}(M_1\rightarrow H_1+\pi)
&=&Ii\varepsilon^{\eta\epsilon^* qv}  g^{p1}_{M_1H_1\pi}\,,\nonumber
\\
\mathcal {M}(M_1\rightarrow S_1+\pi)
&=&I\left[(\eta\cdot q_t)(\epsilon^*\cdot q_t)-\frac{1}{3}(\eta\cdot\epsilon_t^*)q_t^2\right]g^{d2}_{M_1S_1\pi}\,,\nonumber
\\
\mathcal {M}(M_2\rightarrow S_0+\pi)
&=&I\eta_{\alpha_1\alpha_2}\left[q_t^{\alpha_1}q_t^{\alpha_2}-\frac{1}{3}g_t^{\alpha_1\alpha_2}q_t^2\right]g^{d2}_{M_2S_0\pi}\,,\nonumber
\\
\mathcal {M}(M_1\rightarrow T_2+\pi)
&=&2Ii\epsilon^*_{\beta_1\beta_2}\varepsilon^{\beta_1\eta qv}q_t^{\beta_2}g^{d2}_{M_1T_2\pi}\,,\nonumber
\\
\mathcal {M}(M_2\rightarrow T_1+\pi)
&=&2Ii\eta_{\alpha_1\alpha_2}\varepsilon^{\alpha_1\epsilon^* qv}q_t^{\alpha_2}g^{d2}_{M_2T_1\pi}\,,\nonumber
\\
\mathcal {M}(M_2\rightarrow T_2+\pi)
&=&2I\eta_{\alpha_1\alpha_2}\epsilon^*_{\beta_1\beta_2}
\left[g_t^{\alpha_1\beta_1}g_t^{\alpha_2\beta_2}-\frac{1}{3}g_t^{\alpha_1\alpha_2}g_t^{\beta_1\beta_2}\right]g^{s0}_{M_2T_2\pi}\nonumber
\\
&&+I\eta_{\alpha_1\alpha_2}\epsilon^*_{\beta_1\beta_2}
\left\{q_t^{\alpha_1}q_t^{\alpha_2}g_t^{\beta_1\beta_2}+q_t^{\beta_1}q_t^{\beta_2}g_t^{\alpha_1\alpha_2}\right.\nonumber
\\
&&\left.-3q_t^{\alpha_1}q_t^{\beta_1}g_t^{\alpha_2\beta_2}
-q_t^2\left[\frac{2}{3}g_t^{\alpha_1\alpha_2}g_t^{\beta_1\beta_2}-g_t^{\alpha_1\beta_1}g_t^{\alpha_2\beta_2}\right]\right\}g^{d2}_{M_2T_2\pi}\,,\nonumber
\\
\mathcal {M}(M_1\rightarrow M_1+\pi)
&=&Ii\varepsilon^{\eta\epsilon^* qv} g^{p1}_{M_1M_1\pi}\,,\nonumber
\\
\mathcal {M}(M_2\rightarrow M_1+\pi)
&=&2I\eta_{\alpha_1\alpha_2}\left[\epsilon^{*\alpha_1}_tq_t^{\alpha_2}
-\frac{1}{3}g_t^{\alpha_1\alpha_2}(\epsilon^*\cdot q_t)\right]g^{p1}_{M_2M_1\pi}\nonumber
\\
&&+I\eta_{\alpha_1\alpha_2}
\left\{q_t^{\alpha_1}q_t^{\alpha_2}(\epsilon^*\cdot q_t)
-\frac{q_t^2}{5}\left[2\epsilon_t^{*\alpha_1}q_t^{\alpha_2}+(\epsilon^*\cdot q_t)g_t^{\alpha_1\alpha_2}\right]\right\}g^{f3}_{M_2M_1\pi}\,,\nonumber
\\
\mathcal {M}(M_2\rightarrow M_2+\pi)
&=&4Ii\eta_{\alpha_1\alpha_2}\eta^*_{\beta_1\beta_2}\varepsilon^{\alpha_1\beta_1qv}g_t^{\alpha_2\beta_2} g^{p1}_{M_2M_2\pi}\nonumber
\\
&&+4Ii\eta_{\alpha_1\alpha_2}\eta^*_{\beta_1\beta_2}\varepsilon^{\alpha_1\beta_1qv}
\left[q_t^{\alpha_2}q_t^{\beta_2}-\frac{q_t^2}{5}g_t^{\alpha_2\beta_2}\right]g^{f3}_{M_2M_2\pi}.
\end{eqnarray}
We notice that
\begin{eqnarray}
&&g^{p1}_{M_2H_1\pi}=\frac{\sqrt{6}}{2}g^{p1}_{M_1H_0\pi}=\sqrt{6}g^{p1}_{M_1H_1\pi}\,,\nonumber\\
&&g^{d2}_{M_2S_1\pi}=-\frac{\sqrt{6}}{2}g^{d2}_{M_1S_1\pi}=-g^{d2}_{M_2S_0\pi}\,,\nonumber\\
&&g^{s0}_{M_1T_1\pi}=g^{s0}_{M_2T_2\pi}\,,\nonumber\\
&&g^{d2}_{M_1T_1\pi}=-\sqrt{6}g^{d2}_{M_1T_2\pi}=-\sqrt{6}g^{d2}_{M_2T_1\pi}=-\frac{3}{2}g^{d2}_{M_2T_2\pi}\,,\nonumber\\
&&g^{p1}_{M_2M_1\pi}=-\frac{\sqrt{6}}{10}g^{p1}_{M_1M_1\pi}=\frac{\sqrt{6}}{3}g^{p1}_{M_2M_2\pi}\,,\nonumber\\
&&g^{f3}_{M_2M_1\pi}=2\sqrt{6}g^{f3}_{M_2M_2\pi}\,.
\end{eqnarray}
These relations are consistent with the HQET expectation at the
leading order.

To get the numerical values of the six independent coupling constants,
we need the following mass parameters $\bar{\Lambda}$'s and $f$'s, the overlap amplitudes of these
interpolating currents:
\begin{center}
\setlength\extrarowheight{8pt}
\begin{tabular}{ccccc}
\hline
                        &$(0^-,1^-)$ \cite{neubert}          &$(0^+,1^+)$ \cite{dai}               &$(1^+,2^+)$ \cite{dai}              &$(1^-,2^-)$ \cite{pidecaywei}       \\
  $\bar{\Lambda}$\ \ \  & $0.5\ \text{GeV}$                  &$0.85\ \text{GeV}$                   &$0.95\ \text{GeV}$                  &$1.42\ \text{GeV}$                  \\
  $f$\ \ \              & $(0.25\pm 0.04)\ \text{GeV}^{3/2}$ &$(0.36\pm 0.10)\ \text{GeV}^{3/2}$   &$(0.26\pm 0.06)\ \text{GeV}^{5/2}$  &$(0.39\pm 0.03)\ \text{GeV}^{5/2}$  \\
\hline
\end{tabular}
\end{center}
The $\pi$ decay constant $f_\pi=131\ \text{MeV}$. $\mu_\pi\equiv m_\pi^2/(m_u+m_d)$ is given in Ref. \cite{pilcda}:
$\mu_\pi=(1.573\pm 0.174)\ \text{GeV}$.
The parameters appear in the $\pi$ distribution amplitudes are listed below. We use the values at the scale $\mu=1\ \text{GeV}$ in our calculation
under the consideration that the heavy quark behaves almost as a spectator of the decay processes in our discussion at the leading
order of HQET.
\begin{center}
\setlength\extrarowheight{8pt}
\begin{tabular}{ccccccccccc}
\hline
  $a_2$\ \  &$\eta_3$\ \  &$\omega_3$\ \   &$\eta_4$\ \  &$\omega_4$\ \  &$h_{00}$\ \  &$v_{00}$\ \  &$a_{10}$\ \  &$v_{10}$\ \  &$h_{01}$\ \  &$h_{10}$\\
  $0.25$\ \ &$0.015$\ \   &$-1.5$\ \       &$10$\ \      &$0.2$\ \       &$-3.33$\ \   &$-3.33$\ \   &$5.14$\ \    &$5.25$\ \    &$3.46$\ \    &$7.03$  \\
\hline
\end{tabular}
\end{center}

We will work at the symmetry point, i.e., $T_1=T_2=2T$, $u_0=1/2$.
This comes from the consideration that every reliable sum rule has a working interval of the Borel
parameter $T$ within which the sum rule is insensitive to the
variation of $T$. So it is reasonable to choose a common point
$T_1=T_2$ at the overlap of $T_1$ and $T_2$.
Furthermore, choosing $T_1=T_2$ will enable us to subtract the
continuum contribution cleanly, while the asymmetric choice will
lead to the very difficult continuum substraction \cite{asymmetricpoint}.

\begin{figure}
\begin{minipage}[t]{0.5\linewidth}
\centering
\captionstyle{flushleft}
\includegraphics[width=3in]{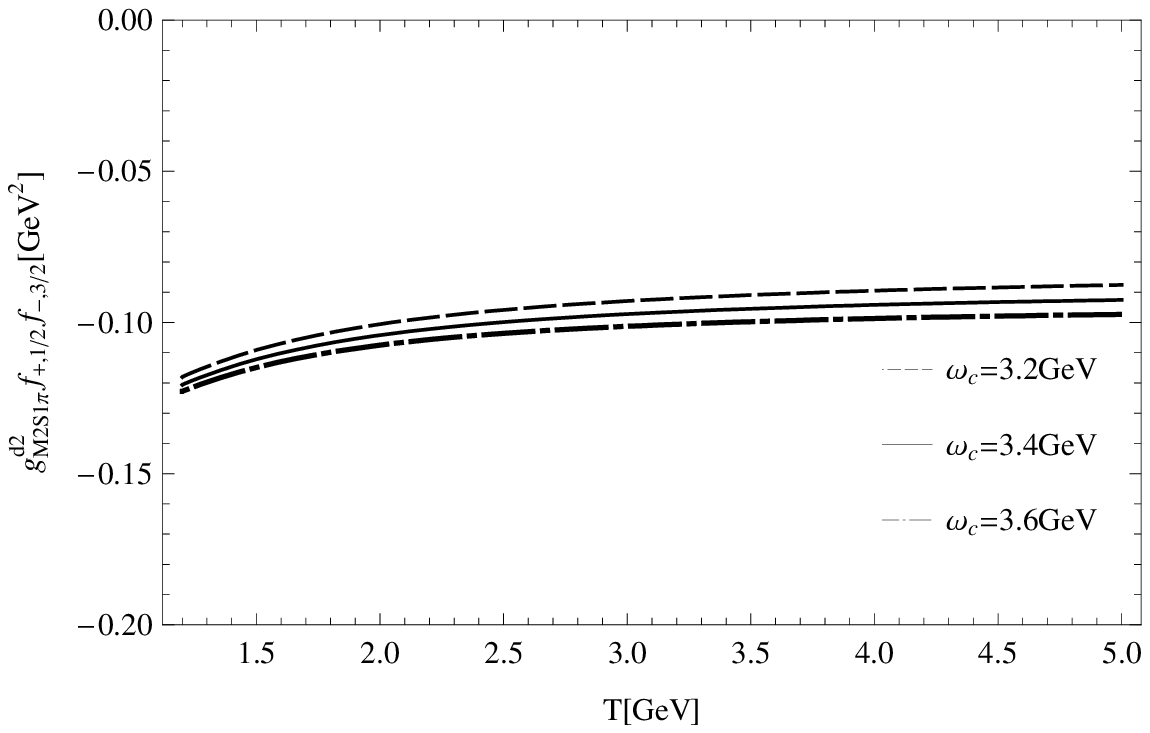}
\setcaptionwidth{3in}
\caption{The sum rule for $g^{d2}_{M_2S_1\pi}f_{+,1/2}f_{-,3/2}$ with $\omega_c=3.2,3.4,3.6\ \text{GeV}$ and the working interval
$3.0<T<4.0\ \text{GeV}$.} \label{fig:CCM2S1D2pi}
\end{minipage}%
\begin{minipage}[t]{0.5\linewidth}
\centering
\captionstyle{flushleft}
\includegraphics[width=3in]{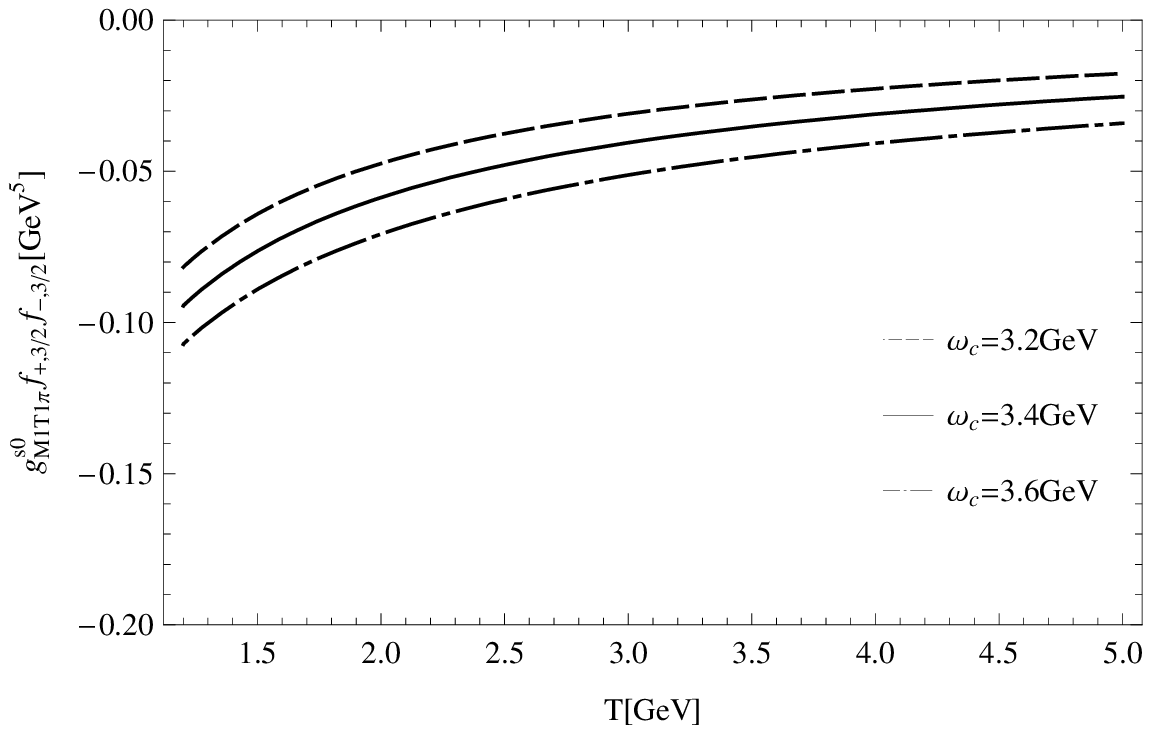}
\setcaptionwidth{3in}
\caption{The sum rule for $g^{s0}_{M_1T_1\pi}f_{+,3/2}f_{-,3/2}$ with $\omega_c=3.2,3.4,3.6\ \text{GeV}$. There is no working
interval for the Borel parameter $T$.} \label{fig:CCM1T1S0pi}
\end{minipage}
\end{figure}

\begin{figure}
\begin{minipage}[t]{0.5\linewidth}
\centering
\captionstyle{flushleft}
\includegraphics[width=3in]{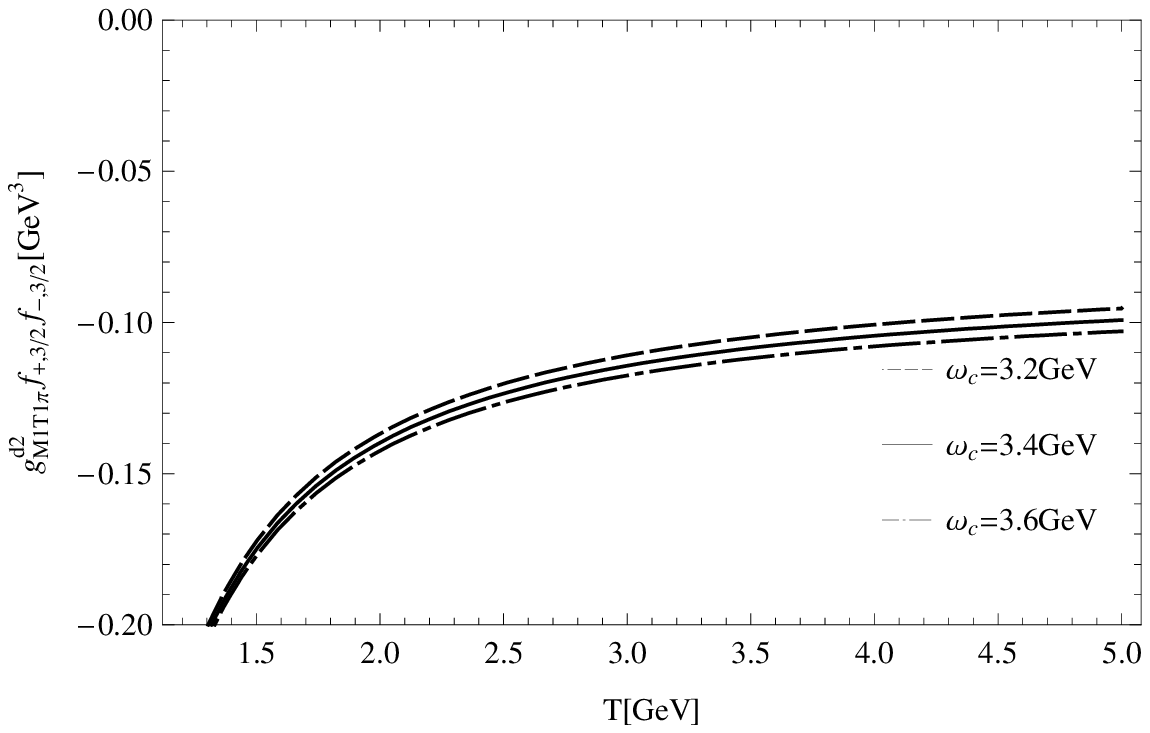}
\setcaptionwidth{3in}
\caption{The sum rule for $g^{d2}_{M_1T_1\pi}f_{+,3/2}f_{-,3/2}$ with $\omega_c=3.2,3.4,3.6\ \text{GeV}$ and the working interval
$3.0<T<4.0\ \text{GeV}$.} \label{fig:CCM1T1D2pi}
\end{minipage}%
\begin{minipage}[t]{0.5\linewidth}
\centering
\captionstyle{flushleft}
\includegraphics[width=3in]{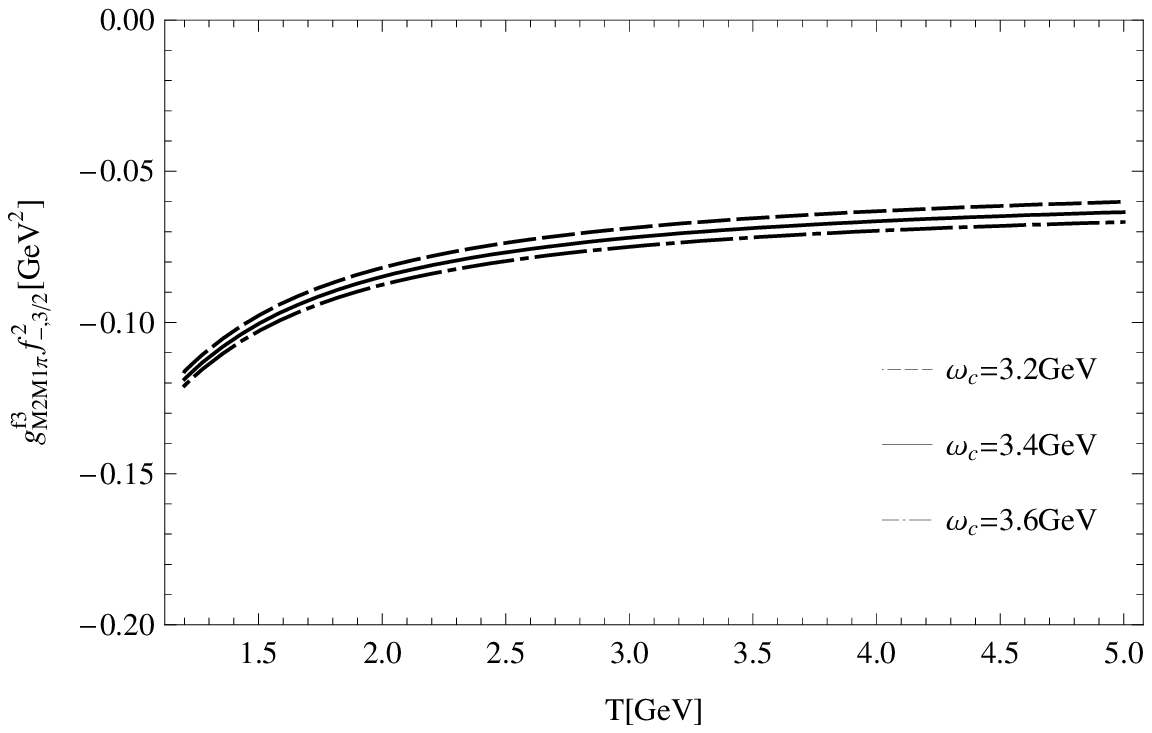}
\setcaptionwidth{3in}
\caption{The sum rule for $g^{f3}_{M_2M_1\pi}f_{-,3/2}^2$ with $\omega_c=3.2,3.4,3.6\ \text{GeV}$ and the working interval
$3.0<T<4.0\ \text{GeV}$.} \label{fig:CCM2M1F3pi}
\end{minipage}
\end{figure}

From the convergence requirement of the
operator product expansion and the requirement that the pole contribution is larger than
$40\%$, we get the working interval of the Borel parameter $T$.
The resulting sum rules are plotted with $\omega_c=3.2,3.4,3.6\ \text{GeV}$ in Figs. \ref{fig:CCM2S1D2pi}-\ref{fig:CCM2M1F3pi}.
The numerical values of these coupling constants are collected in Table \ref{tablepicc}.

\begin{table}[htb]
\begin{center}
\setlength\extrarowheight{8pt}
\begin{tabular}{ccccccc}
\hline
          & $g^{p1*}_{M_2H_1\pi}$ & $g^{d2}_{M_2S_1\pi}$ & $g^{s0*}_{M_1T_1\pi}$ & $g^{d2}_{M_1T_1\pi}$ & $g^{p1*}_{M_2M_1\pi}$ & $g^{f3}_{M_2M_1\pi}$   \\
$g_c$     & $1.13$                & $-0.68$              & $-0.40$               & $-1.09$              & $0.04$                & $-0.46$                \\
$g$       & $0.78\sim1.65$        & $-1.07\sim-0.47$     & $-0.83\sim-0.15$      & $-1.67\sim-0.75$     & $0.03\sim0.05$        & $-0.62\sim-0.34$       \\
\hline
\end{tabular}
\end{center}
\caption{The $\pi$ coupling constants in units of $\text{[GeV]}^{-j}$
with $j$ the orbital angular momentum of the final pion.
$g_c$s correspond to the central values of the overlap amplitudes and
$\tilde{g}$s with $\omega_c=3.4\ \text{GeV}$ and $T=3.5\ \text{GeV}$
where $\tilde{g}^{p1}_{M_2H_1\pi}\equiv g^{p1}_{M_2H_1\pi}f_{-,{1\over 2}}f_{-,{3\over 2}}$ etc.
The ranges of $g$s are determined according to the uncertainty of the overlap amplitudes
and the ranges of $\tilde{g}$s with $3.0<T<4.0\ \text{GeV}$ and $3.2<\omega_c<3.6\ \text{GeV}$.
We use asterisk to indicate the coupling constants from sum rules without a stable working interval.
}
\label{tablepicc}
\end{table}

%\begin{eqnarray}
%\tilde{g}^{p1}_{M_2H_1\pi}=-0.08\pm 0.015\,,
%\tilde{g}^{d2}_{M_2S_1\pi}=-0.25\pm 0.020
%\end{eqnarray}

%%%%%%%%%%%%%%%%%%%%%%%%%%%%%%%%%%%%%%%%%%%%%%%%%%%%%%%%%%%%%%%
\section{Sum Rules for the $\rho$, $\omega$ coupling constants}\label{rhocoupling}
%%%%%%%%%%%%%%%%%%%%%%%%%%%%%%%%%%%%%%%%%%%%%%%%%%%%%%%%%%%%%%%

The sum rules for the $\rho$ ($\omega$) coupling constants can be obtained using the same apporach.
Now the tensor structure of the decay amplitudes are a little more complicated than those of the $\pi$ meson,
due to the spin of the final $\rho$ ($\omega$) meson.
For example, the decay amplitude $\mathcal {M}(M_2\rightarrow H_1+\rho)$ can now be written as
\begin{eqnarray}
\mathcal {M}(M_2\rightarrow H_1+\rho)
&=&
2I\eta_{\alpha_1\alpha_2}\left[-\varepsilon^{\alpha_1 e^*qv}\epsilon^{*\alpha_2}_t
+\frac{1}{3}g_t^{\alpha_1\alpha_2}\varepsilon^{\epsilon^* e^*qv}\right]g^{p1}_{M_2H_1\rho}\nonumber\\
&&
+2I\eta_{\alpha_1\alpha_2}\left[\varepsilon^{\alpha_1\epsilon^* e^*v}q_t^{\alpha_2}
+\varepsilon^{\alpha_1\epsilon^* qv}e_t^{*\alpha_2}\right]g^{p2}_{M_2H_1\rho}\nonumber\\
&&
+2I\eta_{\alpha_1\alpha_2}\biggl\lbrace
\varepsilon^{\alpha_1\epsilon^* qv}q_t^{\alpha_2}(e^*\cdot q_t)
-\frac{q_t^2}{5}\left[\varepsilon^{\alpha_1\epsilon^* q v}e_t^{*\alpha_2}+\varepsilon^{\alpha_1\epsilon^* e^* v}q_t^{\alpha_2}\right]
\biggl\rbrace g^{f2}_{M_2H_1\rho}\,,
\end{eqnarray}
where $e^*$ is the polarization vector of the final $\rho$ meson.

We define other $\rho$ coupling constants as
\begin{eqnarray}
\mathcal {M}(M_1\rightarrow H_0+\rho)
&=&I\epsilon^{\eta e^*qv}g^{p1}_{M_1H_0\rho}\,,\nonumber
\\
\mathcal {M}(M_1\rightarrow H_1+\rho)
&=&Ii\left[(\eta\cdot e_t^*)(\epsilon^*\cdot q_t)-(\eta\cdot q_t)(\epsilon^*\cdot e^*_t)\right]g^{p1}_{M_1H_1\rho}\nonumber\\
&&+Ii\left[(\eta\cdot e_t^*)(\epsilon^*\cdot q_t)+(\eta\cdot q_t)(\epsilon^*\cdot e^*_t)-\frac{2}{3}(\eta\cdot \epsilon^*_t)(e^*\cdot q_t)\right]g^{p2}_{M_1H_1\rho}\nonumber\\
&&+Ii\biggl\lbrace(\eta\cdot q_t)(\epsilon^*\cdot q_t)(e^*\cdot q_t)
-\frac{q_t^2}{5}\left[(\eta\cdot \epsilon^*_t)(e^*\cdot q_t)
+(\eta\cdot e_t^*)(\epsilon^*\cdot q_t)
+(\eta\cdot q_t)(\epsilon^*\cdot e^*_t)\right]\bigg\rbrace g^{f2}_{M_1H_1\rho}\,,\nonumber\\
\mathcal {M}(M_2\rightarrow H_0+\rho)
&=&2Ii\eta_{\alpha_1\alpha_2}\left[e_t^{*\alpha_1}q_t^{\alpha_2}-\frac{1}{3}g_t^{\alpha_1\alpha_2}(e^*\cdot q_t)\right]g^{p2}_{M_2H_0\rho}\nonumber\\
&&+Ii\eta_{\alpha_1\alpha_2}\biggl\lbrace q_t^{\alpha_1}q_t^{\alpha_2}(e^*\cdot q_t)-\frac{q_t^2}{5}
\left[g_t^{\alpha_1\alpha_2}(e^*\cdot q_t)+2e_t^{*\alpha_1}q_t^{\alpha_2}\right]\biggl\rbrace g^{f2}_{M_2H_0\rho}\,,\nonumber
\\
\mathcal {M}(M_1\rightarrow S_0+\rho)
&=&Ii(\eta\cdot e^*_t)g^{s1}_{M_1S_0\rho}+Ii\left[(\eta\cdot q_t)(e^*\cdot q_t)-\frac{1}{3}(\eta\cdot e^*_t)q_t^2\right]g^{d1}_{M_1S_0\rho}\,,\nonumber
\\
\mathcal {M}(M_1\rightarrow S_1+\rho)
&=&I\epsilon^{\eta\epsilon^*e^*v}g^{s1}_{M_1S_1\rho}\nonumber\\
&&+I\left[\epsilon^{\eta\epsilon^*qv}(e^*\cdot q_t)-\frac{1}{3}\epsilon^{\eta\epsilon^*e^*v}q_t^2\right]g^{d1}_{M_1S_1\rho}\nonumber\\
&&+I\left[\epsilon^{\eta e^*qv}(\epsilon^*\cdot q_t)+\epsilon^{\epsilon^*e^*qv}(\eta\cdot q_t)\right]g^{d2}_{M_1S_1\rho}\,,\nonumber
\\
\mathcal {M}(M_2\rightarrow S_0+\rho)
&=&2I\eta_{\alpha_1\alpha_2}\epsilon^{\alpha_1e^*qv}q_t^{\alpha_2}g^{d2}_{M_2S_0\rho}\,,\nonumber
\\
\mathcal {M}(M_2\rightarrow S_1+\rho)
&=&2Ii\eta_{\alpha_1\alpha_2}\left[\epsilon_t^{*\alpha_1}e_t^{*\alpha_2}
-\frac{1}{3}g_t^{\alpha_1\alpha_2}(\epsilon^*\cdot e_t^*)\right] g^{s1}_{M_2S_1\rho}\nonumber\\
&&+2Ii\eta_{\alpha_1\alpha_2}\biggl\lbrace\left[\epsilon_t^{*\alpha_1}q_t^{\alpha_2}
-\frac{1}{3}g_t^{\alpha_1\alpha_2}(\epsilon^*\cdot q_t)\right](e^*\cdot q_t)
-\frac{q_t^2}{3}\left[\epsilon_t^{*\alpha_1}e_t^{*\alpha_2}
-\frac{1}{3}g_t^{\alpha_1\alpha_2}(\epsilon^*\cdot e_t^*)\right]\biggl\rbrace g^{d1}_{M_2S_1\rho}\nonumber\\
&&+2Ii\eta_{\alpha_1\alpha_2}\biggl\lbrace2\left[e_t^{*\alpha_1}q_t^{\alpha_2}(\epsilon^*\cdot q_t)-q_t^{\alpha_1}q_t^{\alpha_2}(\epsilon^*\cdot e_t^*)\right]\nonumber\\
&&+\left[\epsilon^{*\alpha_1}_tq_t^{\alpha_2}-g_t^{\alpha_1\alpha_2}(\epsilon^*\cdot q_t)\right](e^*\cdot q_t)
-\left[\epsilon^{*\alpha_1}_te_t^{*\alpha_2}-g_t^{\alpha_1\alpha_2}(\epsilon^*\cdot e^*_t)\right]q_t^2\biggl\rbrace g^{d2}_{M_2S_1\rho}\,,
\end{eqnarray}

Because of heavy quark symmetry, there are only six independent coupling constants involving these decay modes.
Their sum rules read
\begin{eqnarray}\label{eq:SumRulerho11}
&&g^{p1}_{M_2H_1\rho}f_{-,\frac{1}{2}}f_{-,\frac{3}{2}}e^{-\frac{ \bar{\Lambda }_{-,3/2}+\bar{\Lambda }_{-,1/2}}{T}}\nonumber\\
&&=\frac{1}{64}\biggl\lbrace4f_{\rho }^T\left[(u\varphi_\perp)'(\bar{u}_0)-\varphi_\perp'(\bar{u}_0)\right] T^2f_1(\frac{\omega _c}{T})
  +4f_\rho m_\rho\left[\varphi_\parallel^{[1]}(\bar{u}_0)-g_\perp^{(a)}(\bar{u}_0)
  -2g_\perp^{(v)}(\bar{u}_0)-2g_\perp^{(v)[1]}(\bar{u}_0)\right.\nonumber\\
&&\mathrel{\phantom{=}}\left.+2(ug_\perp^{(v)})(\bar{u}_0)-4\mathcal {A}^{[1]}(u_0)+2\mathcal {V}^{[1,0]}(u_0)\right]Tf_0(\frac{\omega _c}{T})
  +f_\rho^T m_\rho^2\left[A_T'(\bar{u}_0)-(uA_T)'(\bar{u}_0)+16B_T^{[2]}(\bar{u}_0)\right.\nonumber\\
&&\mathrel{\phantom{=}}\left.-8C_T^{[1]}(\bar{u}_0)+8(uC_T)^{[1]}(\bar{u}_0)+8C_T^{[2]}(\bar{u}_0)
  +32\widetilde{S}^{[0,0]}(u_0)+16T_3^{[0,0]}(u_0)-16T_4^{[0,0]}(u_0)\right]\nonumber\\
&&\mathrel{\phantom{=}}-f_\rho m_\rho^3\left[2A^{[1]}(\bar{u}_0)+32\mathcal {A}^{[-1,0]}(u_0)-16\mathcal {T}^{[-1,0]}(u_0)\right]\frac{1}{T}
\biggl\rbrace\,,
\end{eqnarray}
\begin{eqnarray}\label{eq:SumRulerho12}
&&g^{p2}_{M_2H_1\rho}f_{-,\frac{1}{2}}f_{-,\frac{3}{2}}e^{-\frac{ \bar{\Lambda }_{-,3/2}+\bar{\Lambda }_{-,1/2}}{T}}\nonumber\\
&&=\frac{1}{320}\biggl\lbrace12f_{\rho }^T\left[(u\varphi_\perp)'(\bar{u}_0)-\varphi_\perp'(\bar{u}_0)\right] T^2f_1(\frac{\omega _c}{T})
  +8f_\rho m_\rho\left[2(u\varphi_\parallel)(\bar{u}_0)-2\varphi_\parallel(\bar{u}_0)-3\varphi_\parallel^{[1]}(\bar{u}_0)
  -3g_\perp^{(v)}(\bar{u}_0)\right.\nonumber\\
&&\mathrel{\phantom{=}}\left.+3(ug_\perp^{(v)})(\bar{u}_0)+3g_\perp^{(v)[1]}(\bar{u}_0)
  -3\mathcal {V}^{[1,0]}(u_0)\right]Tf_0(\frac{\omega _c}{T})
  +f_\rho^T m_\rho^2\left[32(u\varphi_\perp)^{[1]}(\bar{u}_0)-32\varphi_\perp^{[1]}(\bar{u}_0)+3A_T'(\bar{u}_0)\right.\nonumber\\
&&\mathrel{\phantom{=}}-3(uA_T)'(\bar{u}_0)-32B_T^{[1]}(\bar{u}_0)
  +32(uB_T)^{[1]}(\bar{u}_0)-16B_T^{[2]}(\bar{u}_0)-40C_T^{[1]}(\bar{u}_0)+40(uC_T)^{[1]}(\bar{u}_0)+40C_T^{[2]}(\bar{u}_0)\nonumber\\
&&\mathrel{\phantom{=}}\left.-16\mathcal {T}^{[0,0]}(u_0)-32T_1^{[0,0]}(u_0)-32T_2^{[0,0]}(u_0)
  +48T_3^{[0,0]}(u_0)+48T_4^{[0,0]}(u_0)\right]+f_\rho m_\rho^3\left[64\varphi_\parallel^{[2]}(\bar{u}_0)\right.\nonumber\\
&&\mathrel{\phantom{=}}-64(u\varphi_\parallel)^{[2]}(\bar{u}_0)-64\varphi_\parallel^{[3]}(\bar{u}_0)+4A(\bar{u}_0)-4(uA)(\bar{u}_0)+6A^{[1]}(\bar{u}_0)
  -64g_\perp^{(v)[2]}(\bar{u}_0)+64(ug^{(v)}_\perp)^{[2]}(\bar{u}_0)\nonumber\\
&&\mathrel{\phantom{=}}\left.+64g_\perp^{(v)[3]}(\bar{u}_0)-16\mathcal {V}^{[-1,0]}(u_0)+64\Psi^{[-1,0]}(u_0)\right]\frac{1}{T}
  +8f_\rho^T m_\rho^4\left[A_T^{[1]}(\bar{u}_0)-(uA_T)^{[1]}(\bar{u}_0)+16B_T^{[3]}(\bar{u}_0)\right.\nonumber\\
&&\mathrel{\phantom{=}}-16(uB_T)^{[3]}(\bar{u}_0)-32B_T^{[4]}(\bar{u}_0)
  +8\mathcal {T}^{[-2,0]}(u_0)+16T_1^{[-2,0]}(u_0)+16T_2^{[-2,0]}(u_0)
  +16T_3^{[-2,0]}(u_0)\nonumber\\
&&\mathrel{\phantom{=}}\left.+16T_4^{[-2,0]}(u_0)\right]\frac{1}{T^2}
  +16f_\rho m^5_\rho\left[A^{[3]}(\bar{u}_0)-A^{[2]}(\bar{u}_0)+(uA)^{[2]}(\bar{u}_0)-16\Psi^{[-3,0]}(u_0)\right]\frac{1}{T^3}
\biggl\rbrace\,,
\end{eqnarray}
\begin{eqnarray}\label{eq:SumRulerho13}
&&g^{f2}_{M_2H_1\rho}f_{-,\frac{1}{2}}f_{-,\frac{3}{2}}e^{-\frac{ \bar{\Lambda }_{-,3/2}+\bar{\Lambda }_{-,1/2}}{T}}\nonumber\\
&&=\frac{1}{8}\biggl\lbrace4f_{\rho }^T\left[(u\varphi_\perp)^{[1]}(\bar{u}_0)-\varphi_\perp^{[1]}(\bar{u}_0)\right]
  +8f_\rho m_\rho\left[\varphi_\parallel^{[2]}(\bar{u}_0)-(u\varphi_\parallel)^{[2]}(\bar{u}_0)-\varphi_\parallel^{[3]}(\bar{u}_0)
  -g_\perp^{(v)[2]}(\bar{u}_0)\right.\nonumber\\
&&\mathrel{\phantom{=}}\left.+(ug_\perp^{(v)})^{[2]}(\bar{u}_0)+g_\perp^{(v)[3]}(\bar{u}_0)
  +\mathcal {V}^{[-1,0]}(u_0)\right]\frac{1}{T}+f_\rho^T m_\rho^2\left[A_T^{[1]}(\bar{u}_0)-(uA_T)^{[1]}(\bar{u}_0)
  +16B_T^{[3]}(\bar{u}_0)\right.\nonumber\\
&&\mathrel{\phantom{=}}-16(uB_T)^{[3]}(\bar{u}_0)-32B_T^{[4]}(\bar{u}_0)+8\mathcal {T}^{[-2,0]}(u_0)+16T_1^{[-2,0]}(u_0)+16T_2^{[-2,0]}(u_0)
  +16T_3^{[-2,0]}(u_0)\nonumber\\
&&\mathrel{\phantom{=}}\left.+16T_4^{[-2,0]}(u_0)\right]\frac{1}{T^2}
  +f_\rho m_\rho^3\left[2A^{[3]}(\bar{u}_0)-2A^{[2]}(\bar{u}_0)+2(uA)^{[2]}(\bar{u}_0)-32\Psi^{[-3,0]}(u_0)\right]\frac{1}{T^3}
\biggl\rbrace\,,
\end{eqnarray}
\begin{eqnarray}\label{eq:SumRulerho21}
&&g^{s1}_{M_2S_1\rho}f_{+,\frac{1}{2}}f_{-,\frac{3}{2}}e^{-\frac{ \bar{\Lambda }_{-,3/2}+\bar{\Lambda }_{+,1/2}}{T}}\nonumber\\
&&=\frac{1}{192}\biggl\lbrace4f_{\rho }^T\left[(u\varphi_\perp)''(\bar{u}_0)-\varphi_\perp''(\bar{u}_0)\right]T^3f_2(\frac{\omega _c}{T})
  +f_\rho m_\rho\left[8(u\varphi_\parallel)'(\bar{u}_0)-8\varphi_\parallel'(\bar{u}_0)-16\varphi_\parallel(\bar{u}_0)+16g_\perp^{(v)}(\bar{u}_0)\right.\nonumber\\
&&\mathrel{\phantom{=}}\left.-2(ug_\perp^{(a)})''(\bar{u}_0)+2(g_\perp^{(a)})''(\bar{u}_0)+4(g_\perp^{(a)})'(\bar{u}_0)
  -8\mathcal {A}^{[2,0]}(u_0)+16\mathcal {V}^{[2,0]}(u_0)\right]T^2f_1(\frac{\omega _c}{T})+f_\rho^T m_\rho^2\left[16\varphi_\perp(\bar{u}_0)\right.\nonumber\\
&&\mathrel{\phantom{=}}-16(u\varphi_\perp)(\bar{u}_0)
  -8(uh_\parallel^{(s)})'(\bar{u}_0)+8(h_\parallel^{(s)})'(\bar{u}_0)+24h_\parallel^{(s)}(\bar{u}_0)+A_T''(\bar{u}_0)-(uA_T)''(\bar{u}_0)+16B_T^{[1]}(\bar{u}_0)+24C_T^{[1]}(\bar{u}_0)\nonumber\\
&&\mathrel{\phantom{=}}\left.-8\mathcal {T}^{[1,0]}(u_0)-16T_1^{[1,0]}(u_0)-32T_2^{[1,0]}(u_0)
  +16T_3^{[1,0]}(u_0)-32S^{[1,0]}(u_0)\right]Tf_0(\frac{\omega _c}{T})+f_\rho m_\rho^3\left[32\varphi_\parallel^{[1]}(\bar{u}_0)\right.\nonumber\\
&&\mathrel{\phantom{=}}-32(u\varphi_\parallel)^{[1]}(\bar{u}_0)-32\varphi^{[2]}(\bar{u}_0)
  +32g_\perp^{(v)[2]}(\bar{u}_0)-8g_\perp^{(a)}(\bar{u}_0)+8(ug_\perp^{(a)})(\bar{u}_0)+8g_\perp^{(a)[1]}(\bar{u}_0)+4A(\bar{u}_0)+2A'(\bar{u}_0)\nonumber\\
&&\mathrel{\phantom{=}}-2(uA)'(\bar{u}_0)+16C^{[1]}(\bar{u}_0)-16(uC)^{[1]}(\bar{u}_0)+16C^{[2]}(\bar{u}_0)
  +48\mathcal {V}^{[0,0]}(u_0)-32\mathcal {A}^{[0,0]}(u_0)-32\Psi^{[0,0]}(u_0)\nonumber\\
&&\mathrel{\phantom{=}}\left.-32\widetilde{\Phi}^{[0,0]}(u_0)-32\Phi^{[0,0]}(u_0)\right]+f_\rho^T m_\rho^4\left[32(uh_\parallel^{(s)})^{[1]}(\bar{u}_0)-32h_\parallel^{(s)[1]}(\bar{u}_0)
  -4A_T(\bar{u}_0)+4(uA_T)(\bar{u}_0)\right.\nonumber\\
&&\mathrel{\phantom{=}}\left.-64B_T^{[3]}(\bar{u}_0)-32\mathcal {T}^{[-1,0]}(u_0)-64T_1^{[-1,0]}(u_0)+64T_2^{[-1,0]}(u_0)
  +64T_3^{[-1,0]}(u_0)+64S^{[-1,0]}(u_0)\right]\frac{1}{T}\nonumber\\
&&\mathrel{\phantom{=}}+f_\rho m_\rho^5\left[8A^{[2]}(\bar{u}_0)-8A^{[1]}(\bar{u}_0)+8(uA)^{[1]}(\bar{u}_0)
  -64C^{[3]}(\bar{u}_0)+64(uC)^{[3]}(\bar{u}_0)+128C^{[4]}(\bar{u}_0)+64\mathcal {V}^{[-2,0]}(u_0)\right.\nonumber\\
&&\mathrel{\phantom{=}}\left.+128\Psi^{[-2,0]}(u_0)+128\widetilde{\Phi}^{[-2,0]}(u_0)+128\Phi^{[-2,0]}(u_0)\right]\frac{1}{T^2}
\biggl\rbrace\,,
\end{eqnarray}
\begin{eqnarray}\label{eq:SumRulerho22}
&&g^{d1}_{M_2S_1\rho}f_{+,\frac{1}{2}}f_{-,\frac{3}{2}}e^{-\frac{ \bar{\Lambda }_{-,3/2}+\bar{\Lambda }_{+,1/2}}{T}}\nonumber\\
&&=\frac{1}{32}\biggl\lbrace4f_{\rho }^T\left[(u\varphi_\perp)(\bar{u}_0)-\varphi_\perp(\bar{u}_0)\right]Tf_0(\frac{\omega _c}{T})
  +f_\rho m_\rho\left[16\varphi_\parallel^{[1]}(\bar{u}_0)-16(u\varphi_\parallel)^{[1]}(\bar{u}_0)-16\varphi_\parallel^{[2]}(\bar{u}_0)\right.\nonumber\\
&&\mathrel{\phantom{=}}\left.+16g_\perp^{(v)[2]}(\bar{u}_0)+2g_\perp^{(a)}(\bar{u}_0)-2(ug_\perp^{(a)})(\bar{u}_0)+4g_\perp^{(a)[1]}(\bar{u}_0)
  +8\mathcal {A}^{[0,0]}(u_0)-16\mathcal {V}^{[0,0]}(u_0)\right]\nonumber\\
&&\mathrel{\phantom{=}}+f_\rho^T m_\rho^2\left[16(uh_\parallel^{(s)})^{[1]}(\bar{u}_0)-16h_\parallel^{(s)[1]}(\bar{u}_0)
  +A_T(\bar{u}_0-(uA_T)(\bar{u}_0))-32B_T^{[3]}(\bar{u}_0)-16\mathcal {T}^{[-1,0]}(u_0)\right.\nonumber\\
&&\mathrel{\phantom{=}}\left.-32T_1^{[-1,0]}(u_0)-16T_2^{[-1,0]}(u_0)
  -16T_3^{[-1,0]}(u_0)+32S^{[-1,0]}(u_0)\right]\frac{1}{T}+f_\rho m_\rho^3\left[4A^{[2]}(\bar{u}_0)\right.\nonumber\\
&&\mathrel{\phantom{=}}-4A^{[1]}(\bar{u}_0)+4(uA)^{[1]}(\bar{u}_0)
  -32C^{[3]}(\bar{u}_0)+32(uC)^{[3]}(\bar{u}_0)+64C^{[4]}(\bar{u}_0)+32\mathcal {V}^{[-2,0]}(u_0)\nonumber\\
&&\mathrel{\phantom{=}}\left.+64\Psi^{[-2,0]}(u_0)+64\widetilde{\Phi}^{[-2,0]}(u_0)+64\Phi^{[-2,0]}(u_0)\right]\frac{1}{T^2}
\biggl\rbrace\,,
\end{eqnarray}
\begin{eqnarray}\label{eq:SumRulerho23}
&&g^{d2}_{M_2S_1\rho}f_{+,\frac{1}{2}}f_{-,\frac{3}{2}}e^{-\frac{ \bar{\Lambda }_{-,3/2}+\bar{\Lambda }_{+,1/2}}{T}}\nonumber\\
&&=\frac{1}{32}\biggl\lbrace4f_{\rho }^T\left[\varphi_\perp(\bar{u}_0)-(u\varphi_\perp)(\bar{u}_0)\right]Tf_0(\frac{\omega _c}{T})
  +2f_\rho m_\rho\left[(ug_\perp^{(a)})(\bar{u}_0)-g_\perp^{(a)}(\bar{u}_0)+4\mathcal {A}^{[0,0]}(u_0)\right]\nonumber\\
&&\mathrel{\phantom{=}}+f_\rho^T m_\rho^2\left[(uA_T)(\bar{u}_0)-A_T(\bar{u}_0)+16T_2^{[-1,0]}(u_0)+16T_3^{[-1,0]}(u_0)\right]\frac{1}{T}
\biggl\rbrace\,.
\end{eqnarray}

The other $\rho$ coupling constants are related to the above ones by the following relations:
\begin{eqnarray}
&&g^{p1}_{M_2H_1\rho}=\frac{\sqrt{6}}{4}g^{p1}_{M_1H_0\rho}=-\frac{\sqrt{6}}{2}g^{p1}_{M_1H_1\rho}\,,\nonumber\\
&&g^{p2}_{M_2H_1\rho}=\frac{\sqrt{6}}{6}g^{p2}_{M_1H_1\rho}=-\frac{1}{2}g^{p2}_{M_2H_0\rho}\,,\nonumber\\
&&g^{f2}_{M_2H_1\rho}=\frac{\sqrt{6}}{6}g^{f2}_{M_1H_1\rho}=-\frac{1}{2}g^{f2}_{M_2H_0\rho}\,,\nonumber\\
&&g^{s1}_{M_2S_1\rho}=-\frac{\sqrt{6}}{4}g^{s1}_{M_1S_0\rho}=-\frac{\sqrt{6}}{2}g^{s1}_{M_1S_1\rho}\,,\nonumber\\
&&g^{d1}_{M_2S_1\rho}=-\frac{\sqrt{6}}{4}g^{d1}_{M_1S_0\rho}=\frac{\sqrt{6}}{2}g^{d1}_{M_1S_1\rho}\,,\nonumber\\
&&g^{d2}_{M_2S_1\rho}=-\frac{\sqrt{6}}{6}g^{d2}_{M_1S_1\rho}=-\frac{1}{2}g^{d2}_{M_2S_0\rho}\,.
\end{eqnarray}

In our numerical analysis, the parameters that appear in the distribution amplitudes of the $\rho$ meson
take the values from Ref. \cite{rhoparameter}.
\begin{center}
\setlength\extrarowheight{8pt}
\begin{tabular}{cccccccccccc}
\hline
  $f_\rho$[MeV]&$f^T_\rho$[MeV]&$a^\parallel_2$&$a^\perp_2$&$\zeta^\parallel_{3\rho}$
  &$\widetilde{\omega}^\parallel_{3\rho}$&$\omega^\parallel_{3\rho}$&$\omega^\perp_{3\rho}$&$\zeta^\parallel_4$
  &$\widetilde{\omega}^\parallel_4$&$\zeta^\perp_4$&$\widetilde{\zeta}^\perp_4$\\
  $216(3)$&$165(9)$&$0.15(7)$&$0.14(6)$&$0.030(10)$&$-0.09(3)$&$0.15(5)$&$0.55(25)$
  &$0.07(3)$&$-0.03(1)$&$-0.03(5)$&$-0.08(5)$  \\
\hline
\end{tabular}
\end{center}

\begin{table}[htb]
\begin{center}
\setlength\extrarowheight{8pt}
\begin{tabular}{ccccccc}
\hline
          & $g^{p1*}_{M_2H_1\rho}$ & $g^{p2*}_{M_2H_1\rho}$ & $g^{f2}_{M_2H_1\rho}$ & $g^{s1*}_{M_2S_1\rho}$ & $g^{d1}_{M_2S_1\rho}$ & $g^{d2}_{M_2S_1\rho}$   \\
$g_c$     & $-0.10$                & $-0.26$              & $-0.62$                 & $0.16$                  & $0.32$                & $0.30$                \\
$g$       & $-0.20\sim-0.04$       & $-0.40\sim-0.16$     & $-0.93\sim-0.41$        & $0.09\sim0.27$          & $0.21\sim0.53$        & $0.21\sim0.47$       \\
\hline\hline
          & $g^{p1*}_{M_2H_1\omega}$ & $g^{p2*}_{M_2H_1\omega}$ & $g^{f2}_{M_2H_1\omega}$ & $g^{s1*}_{M_2S_1\omega}$ & $g^{d1}_{M_2S_1\omega}$ & $g^{d2}_{M_2S_1\omega}$   \\
$g_c$     & $-0.08$                & $-0.24$              & $-0.56$                 & $0.14$                  & $0.32$                & $0.27$                \\
$g$       & $-0.13\sim-0.04$       & $-0.34\sim-0.16$     & $-0.79\sim-0.41$        & $0.08\sim0.25$          & $0.21\sim0.53$        & $0.19\sim0.43$       \\
\hline
\end{tabular}
\end{center}
\caption{The $\rho/\omega$ coupling constants in units of $\text{[GeV]}^{-j}$
with $j$ the orbital angular momentum of the final $\rho/\omega$ meson.
$g_c$s correspond to the central values of the overlap amplitudes and
$\tilde{g}$s with $\omega_c=3.4\ \text{GeV}$ and $T=2.75\ \text{GeV}$.
The ranges of $g$s are determined according to the uncertainty of the overlap amplitudes
and the ranges of $\tilde{g}$s with $2.5<T<3.0\ \text{GeV}$ and $3.2<\omega_c<3.6\ \text{GeV}$.
}
\label{tablerhoomegacc}
\end{table}

\begin{figure}[htbp]
\begin{minipage}[t]{0.5\linewidth}
\centering
\captionstyle{flushleft}
\includegraphics[width=3in]{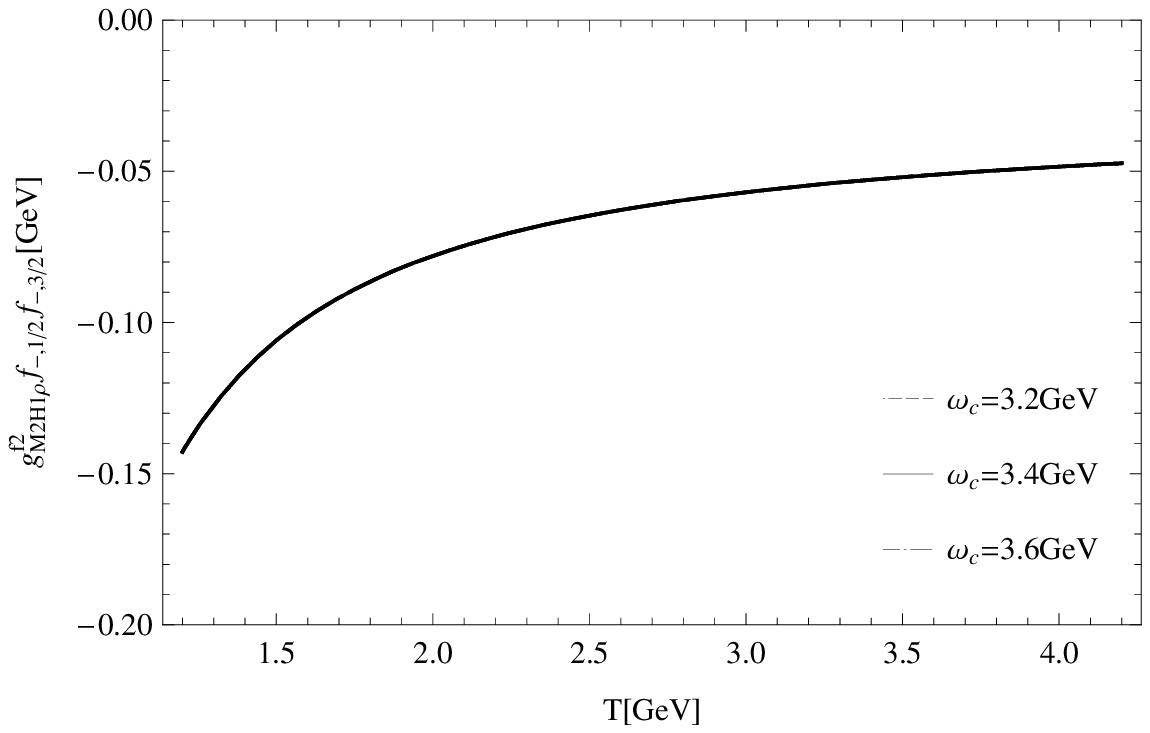}
\setcaptionwidth{3in}
\caption{The sum rule for $g^{f2}_{M_2H_1\rho}f_{-,1/2}f_{-,3/2}$ with $\omega_c=3.2,3.4,3.6\ \text{GeV}$ and the working interval
$2.5<T<3.0\ \text{GeV}$.} \label{fig:CCM2H1F2rho}
\end{minipage}%
\begin{minipage}[t]{0.5\linewidth}
\centering
\captionstyle{flushleft}
\includegraphics[width=3in]{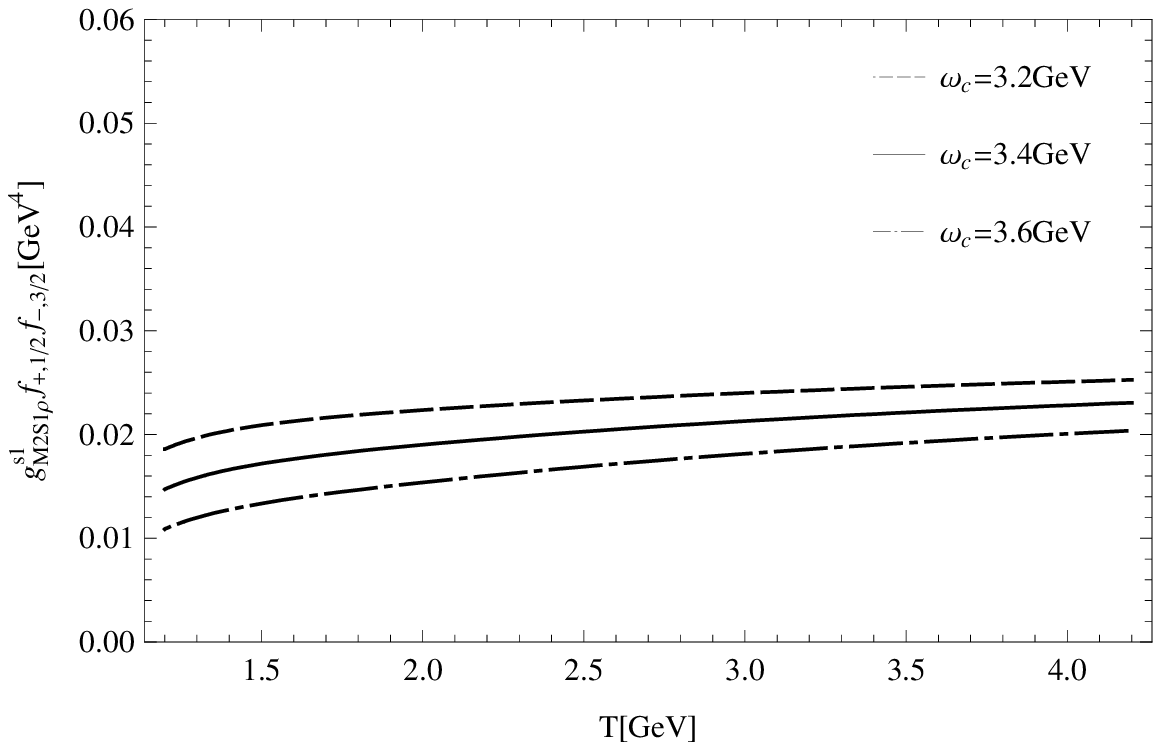}
\setcaptionwidth{3in}
\caption{The sum rule for $g^{s1}_{M_2S_1\rho}f_{+,1/2}f_{-,3/2}$ with $\omega_c=3.2,3.4,3.6\ \text{GeV}$. There is no working interval
for the Borel parameter $T$.} \label{fig:CCM2S1S1rho}
\end{minipage}
\end{figure}

\begin{figure}[htbp]
\begin{minipage}[t]{0.5\linewidth}
\centering
\captionstyle{flushleft}
\includegraphics[width=3in]{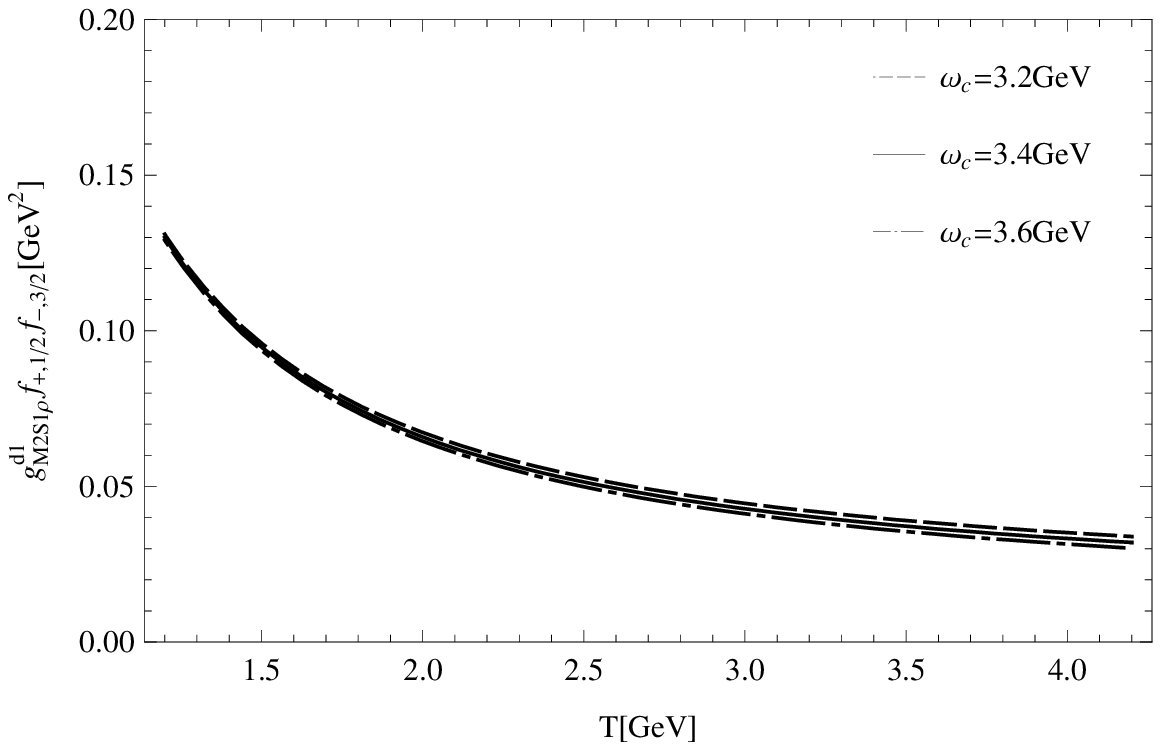}
\setcaptionwidth{3in}
\caption{The sum rule for $g^{d1}_{M_2S_1\rho}f_{+,1/2}f_{-,3/2}$ with $\omega_c=3.2,3.4,3.6\ \text{GeV}$ and the working interval
$2.5<T<3.0\ \text{GeV}$.} \label{fig:CCM2S1D1rho}
\end{minipage}%
\begin{minipage}[t]{0.5\linewidth}
\centering
\captionstyle{flushleft}
\includegraphics[width=3in]{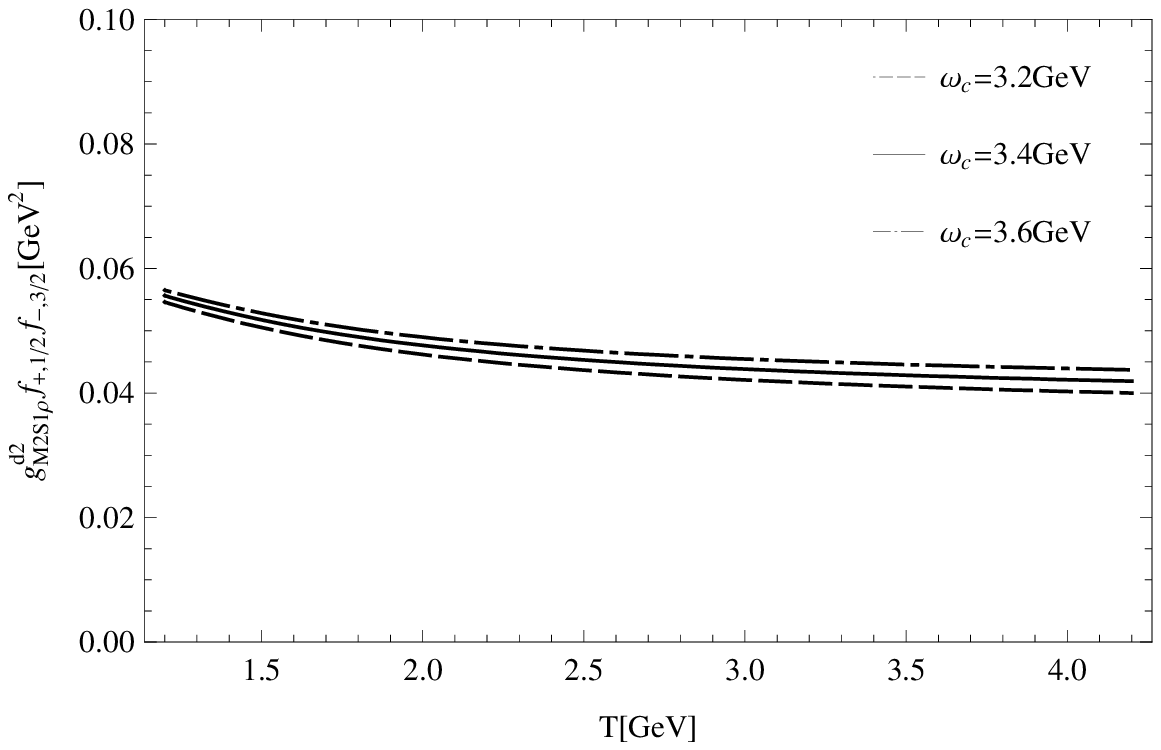}
\setcaptionwidth{3in}
\caption{The sum rule for $g^{d2}_{M_2S_1\rho}f_{+,1/2}f_{-,3/2}$ with $\omega_c=3.2,3.4,3.6\ \text{GeV}$ and the working interval
$2.5<T<3.0\ \text{GeV}$.} \label{fig:CCM2S1D2rho}
\end{minipage}
\end{figure}

The resulting sum rules are plotted with $\omega_c=3.2,3.4,3.6\ \text{GeV}$ in Figs. \ref{fig:CCM2H1F2rho}-\ref{fig:CCM2S1D2rho}.
The numerical values of these coupling constants are collected in Table \ref{tablerhoomegacc}.

\iffalse
\begin{align}
g^{p1}_{M_2H_1\rho}&=-0.10\ \text{GeV}^{-1}\ *\,,         &&  g^{s1}_{M_2S_1\rho}=0.14\ *,                          \nonumber\\
g^{p2}_{M_2H_1\rho}&=-0.20\ \text{GeV}^{-1}\ *\,,         &&  g^{d1}_{M_2S_1\rho}=(0.25\pm 0.04)\ \text{GeV}^{-2},   \nonumber\\
g^{f2}_{M_2H_1\rho}&=(-0.53\pm 0.02)\ \text{GeV}^{-3}\,,  &&  g^{d2}_{M_2S_1\rho}=(0.24\pm 0.02)\ \text{GeV}^{-2}.
\end{align}
The errors come from the variations of $T$ and $\omega_c$ in the working interval
and the central value corresponds to $T=2.75\ \text{GeV}$ and $\omega_c=3.4\ \text{GeV}$.

\begin{table}[htb]
\begin{center}
\setlength\extrarowheight{8pt}
\begin{tabular}{cccccc}
\hline
 $g^{p1}_{M_2H_1\rho}$       &       $g^{p2}_{M_2H_1\rho}$       &      $g^{f2}_{M_2H_1\rho}$          \\
$-0.10\ \text{GeV}^{-1}\ *$  & $-0.26\ \text{GeV}^{-1}\ *$       & $(-0.58\pm 0.04)\ \text{GeV}^{-3}$  \\\hline
 $g^{s1}_{M_2S_1\rho}$       &       $g^{d1}_{M_2S_1\rho}$       &      $g^{d2}_{M_2S_1\rho}$          \\
$-0.19\ *$                   & $(0.38\pm 0.06)\ \text{GeV}^{-2}$ & $(0.29\pm 0.03)\ \text{GeV}^{-2}$   \\\hline
\end{tabular}
\end{center}
\caption{The $\rho$ coupling constants. Here asterisk indicates value from sum rule without a stable working interval.
The errors come from the variations of $T$ and $\omega_c$ in the working interval
and the central value corresponds to $T=2.75\ \text{GeV}$ and $\omega_c=3.4\ \text{GeV}$.}
\label{tablerho}
\end{table}
\fi

%\begin{eqnarray}
%\tilde{g}^{f2}_{M_2H_1\rho}=-0.057\pm 0.004\,,\nonumber\\
%\tilde{g}^{d1}_{M_2S_1\rho}=-0.060\pm 0.010\,,\nonumber\\
%\tilde{g}^{d2}_{M_2S_1\rho}=-0.045\pm 0.005\,.
%\end{eqnarray}

Replacing the $\rho$ meson parameters by those for the $\omega$
meson, one obtains the $\omega$ meson couplings with the heavy
mesons, which are also listed in Table \ref{tablerhoomegacc}.
Here we take the following values for the parameters $f_\omega$, $f_\omega^T$ \cite{rholcda}, and $m_\omega$:
$f_\omega=0.195\ \mbox{GeV}$, $f_\omega^T=0.145\ \mbox{GeV}$, and $m_\omega=0.78\ \mbox{GeV}$.

%\begin{eqnarray}
%\tilde{g}^{f2}_{M_2H_1\omega}=-0.050\pm 0.005\,,\nonumber\\
%\tilde{g}^{d1}_{M_2S_1\omega}=-0.050\pm 0.010\,,\nonumber\\
%\tilde{g}^{d2}_{M_2S_1\omega}=-0.040\pm 0.005\,.
%\end{eqnarray}

%%%%%%%%%%%%%%%%%%%%%%%%%%%%%%%%%%%%%%%%%%%%%%%%%%%%%%%%%%%%%%%
\section{Sum Rules for the $\gamma$ coupling constants}\label{gammacoupling}
%%%%%%%%%%%%%%%%%%%%%%%%%%%%%%%%%%%%%%%%%%%%%%%%%%%%%%%%%%%%%%%

As an example, we consider the decay $M_1\rightarrow H_1+\gamma$.
The decay amplitude should now be expressed according to
the total angular momentum of the final photon, namely $m1,e2\cdots$:
\begin{eqnarray}
\mathcal {M}(M_1\rightarrow H_1+\gamma)
&=&ei\left[(\eta\cdot e^*_t)(\epsilon^*\cdot q_t)-(\eta\cdot q_t)(\epsilon^*\cdot e^*_t)\right]g^{m1}_{M_1H_1\gamma}\nonumber\\
&&+ei\left\{(\eta\cdot q_t)(\epsilon^*\cdot q_t)(e^*\cdot q_t)
-\frac{q_t^2}{2}\left[(\eta\cdot e^*_t)(\epsilon^*\cdot q_t)+(\eta\cdot q_t)(\epsilon^*\cdot e^*_t)\right]\right\} g^{e2}_{M_1H_1\gamma}\,,
\end{eqnarray}
where $e$ is the charge of the proton.
To extract $g^{m1}_{M_1H_1\gamma}$ and $g^{e2}_{M_1H_1\gamma}$, we consider the following correlator:
\begin{eqnarray}
\int d^4 xe^{-ik\cdot x}\langle \gamma(q)|T\{J^\beta_{1,-,\frac{1}{2}}(0)J^{\dag\alpha}_{1,-,\frac{3}{2}}(x)\}|0\rangle
&=&ei\left[e^{*\alpha}_tq^\beta_t-e^{*\beta}_tq^\alpha_t\right]G^{m1}_{M_1H_1\gamma}(\omega,\omega')\nonumber\\
&&+ei\biggl\lbrace q_t^\alpha q_t^\beta(e^*\cdot q_t)
-\frac{q_t^2}{2}\left[e^{*\alpha}_tq^\beta_t+e^{*\beta}_tq^\alpha_t\right]\biggl\rbrace G^{e2}_{M_1H_1\gamma}(\omega,\omega')\,.
\end{eqnarray}
In the limit $m_Q\rightarrow \infty$, two diagrams contribute to
the above correlator according to the way how the photon couples
to the light quark. First, the photon couples to the light quark
perturbatively through the standard QED interaction. For this
diagram, we need the quark propagator in an external
electromagnetic field:
\begin{equation}
\langle 0|T\{q^a(x){\bar q}^b(0)\}|0\rangle_{F_{\mu\nu}}
=\frac{\delta^{ab}e_qe}{16\pi^2x^2}\int_0^1
du\{2(1-2u)x_\mu\gamma_\nu
+i\varepsilon_{\mu\nu\rho\sigma}\gamma_5\gamma^\rho x^\sigma
\}F^{\mu\nu}(ux)\,,
\end{equation}
where we have adopted the Fock-Schwinger gauge $x^\mu A_\mu(x)=0$
to express the electromagnetic vector potential in terms of the
gauge invariant $F_{\mu\nu}$. In our calculation, we take the $u$
quark as the light quark involved and therefore
$e_q=\frac{2}{3}$. The second diagram involves the nonperturbative interaction of the
photon with the light quark in terms of the photon light-cone
distribution amplitudes.

Now the problem can be tackled using the same approach as in
Section II and III. We list the final sum rules below:
\begin{eqnarray}\label{eq:SumRulegamma1}
&&g^{m1}_{M_1H_1\gamma}f_{-,\frac{1}{2}}f_{-,\frac{3}{2}}e^{-\frac{\bar{\Lambda }_{-,3/2}+\bar{\Lambda }_{-,1/2}}{T}}\nonumber\\
&&=
-\frac{1}{32\sqrt{6}}e_q\biggl\lbrace\frac{1}{\pi^2}T^3f_2(\frac{\omega_c}{T})
-4\chi\langle\bar{q}q\rangle\left[\phi_\gamma'(\bar{u}_0)-(u\phi_\gamma)'(\bar{u}_0)\right]T^2f_1(\frac{\omega_c}{T})
-4f_{3\gamma}\left[\psi^a(\bar{u}_0)+2\psi^v(\bar{u}_0)\right.\nonumber\\
&&\mathrel{\phantom{=}}\left.
-2(u\psi^v)(\bar{u}_0)+4\mathcal {A}^{[1,0]}(u_0)+2\mathcal {V}^{[1,0]}(u_0)\right]Tf_0(\frac{\omega_c}{T})
+\langle\bar{q}q\rangle\left[A'(\bar{u}_0)-(uA)'(\bar{u}_0)+8h_\gamma^{[2]}(\bar{u}_0)-8h_\gamma^{[1]}(\bar{u}_0)\right.\nonumber\\
&&\mathrel{\phantom{=}}\left.
+8(uh_\gamma)^{[1]}(\bar{u}_0)+32\widetilde{S}^{[0,0]}(u_0)+16T_3^{[0,0]}(u_0)-16T_4^{[0,0]}(u_0)\right]\biggl\rbrace\,,\\
&&
g^{e2}_{M_1H_1\gamma}f_{-,\frac{1}{2}}f_{-,\frac{3}{2}}e^{-\frac{\bar{\Lambda }_{-,3/2}+\bar{\Lambda }_{-,1/2}}{T}}\nonumber\\
&&=
-\frac{1}{8\sqrt{6}}e_q\biggl\lbrace\frac{4u_0^3-3u_0^2}{\pi^2}Tf_0(\frac{\omega_c}{T})
+24\chi\langle\bar{q}q\rangle\left[\phi_\gamma^{[1]}(\bar{u}_0)-(u\phi_\gamma)^{[1]}(\bar{u}_0)\right]
+48f_{3\gamma}\left[\psi^{v[2]}(\bar{u}_0)-(u\psi^v)^{[2]}(\bar{u}_0)\right.\nonumber\\
&&\mathrel{\phantom{=}}\left.
+\mathcal {V}^{[-1,0]}(u_0)\right]\frac{1}{T}-6\langle\bar{q}q\rangle\left[A^{[1]}(\bar{u}_0)-(uA)^{[1]}(\bar{u}_0)
+8h_\gamma^{[4]}(\bar{u}_0)-8h_\gamma^{[3]}(\bar{u}_0)+8(uh_\gamma)^{[3]}(\bar{u}_0)-16T_3^{[-2,0]}(u_0)\right.\nonumber\\
&&\mathrel{\phantom{=}}\left.
-16T_4^{[-2,0]}(u_0)\right]\frac{1}{T^2}\biggl\rbrace\,.
\end{eqnarray}

Similarly, the electromagnetic coupling constants between doublets $M$ and $S$ can be defined as
\begin{eqnarray}
\mathcal {M}(M_1\rightarrow S_1+\gamma)
&=&e\left[\varepsilon^{\eta\epsilon^*e^*v}q_t^2-\varepsilon^{\eta\epsilon^*qv}(e^*\cdot q_t)\right]g^{e1}_{M_1S_1\gamma}\nonumber\\
&&+e\left[2\varepsilon^{\eta e^*qv}(\epsilon^*\cdot q_t)+\varepsilon^{\eta\epsilon^*e^*v}q^2_t-\varepsilon^{\eta\epsilon^*qv}(e^*\cdot q_t)\right]g^{m2}_{M_1S_1\gamma}\,,
\end{eqnarray}
and the sum rules for $g^{e1}_{M_1S_1\gamma}$ and $g^{m2}_{M_1S_1\gamma}$ read
\begin{eqnarray}\label{eq:SumRulegamma2}
&&g^{e1}_{M_1S_1\gamma}f_{+,\frac{1}{2}}f_{-,\frac{3}{2}}e^{-\frac{\bar{\Lambda }_{-,3/2}+\bar{\Lambda }_{+,1/2}}{T}}\nonumber\\
&&=
\frac{1}{16\sqrt{6}}e_q\biggl\lbrace\frac{2u_0^2-u_0}{\pi^2}T^2f_1(\frac{\omega_c}{T})
-4\chi\langle\bar{q}q\rangle\left[\phi_\gamma(\bar{u}_0)-(u\phi_\gamma)(\bar{u}_0)\right]Tf_0(\frac{\omega_c}{T})
+2f_{3\gamma}\left[4\psi^a(\bar{u}_0)+2\psi^{a[1]}(\bar{u}_0)\right.\nonumber\\
&&\mathrel{\phantom{=}}\left.
-(u\psi^a)(\bar{u}_0)+\mathcal {A}^{[0,0]}(u_0)+8\mathcal {V}^{[0,0]}(u_0)\right]
+\langle\bar{q}q\rangle\left[A(\bar{u}_0)-(uA)(\bar{u}_0)+16h_\gamma^{[3]}(\bar{u}_0)
+32\widetilde{S}^{[-1,0]}(u_0)\right.\nonumber\\
&&\mathrel{\phantom{=}}\left.
-16T_1^{[-1,0]}(u_0)+16T_4^{[-1,0]}(u_0)\right]\biggl\rbrace\,,\\
&&
g^{m2}_{M_1S_1\gamma}f_{+,\frac{1}{2}}f_{-,\frac{3}{2}}e^{-\frac{\bar{\Lambda }_{-,3/2}+\bar{\Lambda }_{+,1/2}}{T}}\nonumber\\
&&=
-\frac{\sqrt{6}}{32}e_q\biggl\lbrace\frac{u_0}{\pi^2}T^2f_1(\frac{\omega_c}{T})
+4\chi\langle\bar{q}q\rangle\left[\phi_\gamma(\bar{u}_0)-(u\phi_\gamma)(\bar{u}_0)\right]Tf_0(\frac{\omega_c}{T})
+2f_{3\gamma}\left[4\mathcal {A}^{[0,0]}(u_0)-\psi^a(\bar{u}_0)+(u\psi^a)(\bar{u}_0)\right]\nonumber\\
&&\mathrel{\phantom{=}}
-\langle\bar{q}q\rangle\left[A(\bar{u}_0)-(uA)(\bar{u}_0)+16T_1^{[-1,0]}(u_0)+16T_4^{[-1,0]}(u_0)\right]\biggl\rbrace\,.
\end{eqnarray}

The $\gamma$ coupling constants of the other channels are defined as
\begin{eqnarray}
\mathcal {M}(M_1\rightarrow H_0+\gamma)
&=&e\varepsilon^{\eta e^*qv}g^{m1}_{M_1H_0\gamma}\,,\nonumber
\\
\mathcal {M}(M_2\rightarrow H_0+\gamma)
&=&ei\eta_{\alpha_1\alpha_2}\left[q_t^{\alpha_1}q_t^{\alpha_2}(e^*\cdot q_t)
-e_t^{*\alpha_1}q_t^{\alpha_2}q^2_t\right]g^{e2}_{M_2H_0\gamma}\,,\nonumber
\\
\mathcal {M}(M_2\rightarrow H_1+\gamma)
&=&2e\eta_{\alpha_1\alpha_2}\left[\varepsilon^{\alpha_1\epsilon^*qv}e^{*\alpha_2}_t-\varepsilon^{\alpha_1\epsilon^*e^*v}q^{\alpha_2}_t
  +\frac{2}{3}g^{\alpha_1\alpha_2}_t\varepsilon^{\epsilon^*e^*qv}\right]g^{m1}_{M_2H_1\gamma}\nonumber\\
&&+e\eta_{\alpha_1\alpha_2}\biggl\lbrace q_t^2
\left[\varepsilon^{\alpha_1\epsilon^*qv}e^{*\alpha_2}_t
+\varepsilon^{\alpha_1\epsilon^*e^*v}q^{\alpha_2}_t\right]-2\varepsilon^{\alpha_1\epsilon^*qv}q^{\alpha_2}_t(e^*\cdot q_t)\biggl\rbrace g^{e2}_{M_2H_1\gamma}\,,\nonumber
\\
\mathcal {M}(M_1\rightarrow S_0+\gamma)
&=&ei\left[(\eta\cdot q_t)(e^*\cdot q_t)-(\eta\cdot e^*_t)q^2_t\right]g^{e1}_{M_1S_0\gamma}\,,\nonumber
\\
\mathcal {M}(M_2\rightarrow S_0+\gamma)
&=&2e\eta_{\alpha_1\alpha_2}\varepsilon^{\alpha_1e^*qv}q^{\alpha_2}_tg^{m2}_{M_2S_0\gamma}\,,\nonumber
\end{eqnarray}
\begin{eqnarray}
&&\mathcal {M}(M_2\rightarrow S_1+\gamma)\nonumber\\
&&=2ei\eta_{\alpha_1\alpha_2}\biggl\lbrace
\left[\epsilon^{*\alpha_1}_tq^{\alpha_2}_t
-\frac{1}{3}g^{\alpha_1\alpha_2}_t(\epsilon^*\cdot q_t)\right](e^*\cdot q_t)-\left[\epsilon^{*\alpha_1}_te^{*\alpha_2}_t
-\frac{1}{3}g^{\alpha_1\alpha_2}_t(\epsilon^*\cdot e^*_t)\right]q^2_t\biggl\rbrace g^{e1}_{M_2S_1\gamma}\nonumber\\
&&\mathrel{\phantom{=}}+2ei\eta_{\alpha_1\alpha_2}\biggl\lbrace
2\left[e_t^{*\alpha_1}q_t^{\alpha_2}(\epsilon^*\cdot q_t)-q_t^{\alpha_1}q_t^{\alpha_2}(\epsilon^*\cdot e^*_t)\right]
+\left[\epsilon_t^{*\alpha_1}q_t^{\alpha_2}-g_t^{\alpha_1\alpha_2}(\epsilon^*\cdot q_t)\right](e^*\cdot q_t)\nonumber\\
&&\mathrel{\phantom{=}}-\left[\epsilon_t^{*\alpha_1}e_t^{*\alpha_2}-g_t^{\alpha_1\alpha_2}(\epsilon^*\cdot e^*_t)\right]q_t^2
\biggl\rbrace g^{m2}_{M_2S_1\gamma}\,.
\end{eqnarray}
There are simple relations among the coupling constants of a certain decay type ($e1, m1\cdots$) within two doublets:
\begin{eqnarray}
&&g^{m1}_{M_1H_1\gamma}=-\frac{1}{2}g^{m1}_{M_1H_0\gamma}=-\frac{\sqrt{6}}{3}g^{m1}_{M_2H_1\gamma}\,,\nonumber\\
&&g^{e2}_{M_1H_1\gamma}=-\frac{\sqrt{6}}{2}g^{e2}_{M_2H_0\gamma}=\sqrt{6}g^{e2}_{M_2H_1\gamma}\,,\nonumber\\
&&g^{e1}_{M_1S_1\gamma}=-\frac{1}{2}g^{e1}_{M_1S_0\gamma}=\frac{\sqrt{6}}{3}g^{e1}_{M_2S_1\gamma}\,,\nonumber\\
&&g^{m2}_{M_1S_1\gamma}=\frac{\sqrt{6}}{2}g^{m2}_{M_2S_0\gamma}=-\sqrt{6}g^{m2}_{M_2S_1\gamma}\,.
\end{eqnarray}

\begin{figure}
\begin{minipage}[t]{0.5\linewidth}
\centering
\captionstyle{flushleft}
\includegraphics[width=3in]{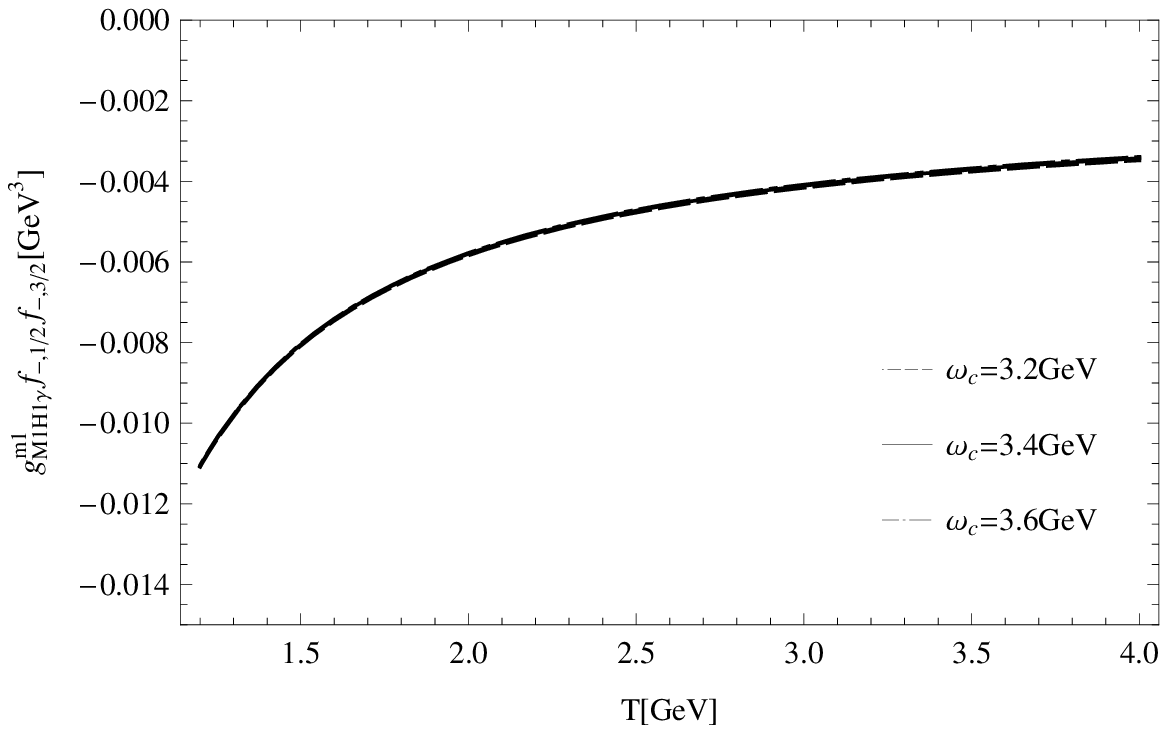}
\setcaptionwidth{3in}
\caption{The sum rule for $g^{m1}_{M_1H_1\gamma}f_{-,1/2}f_{-,3/2}$ with $\omega_c=3.2,3.4,3.6\ \text{GeV}$ and the working interval
$2.5<T<3.5\ \text{GeV}$.}\label{fig:CCM1H1M1gamma}
\end{minipage}%
\begin{minipage}[t]{0.5\linewidth}
\centering
\captionstyle{flushleft}
\includegraphics[width=3in]{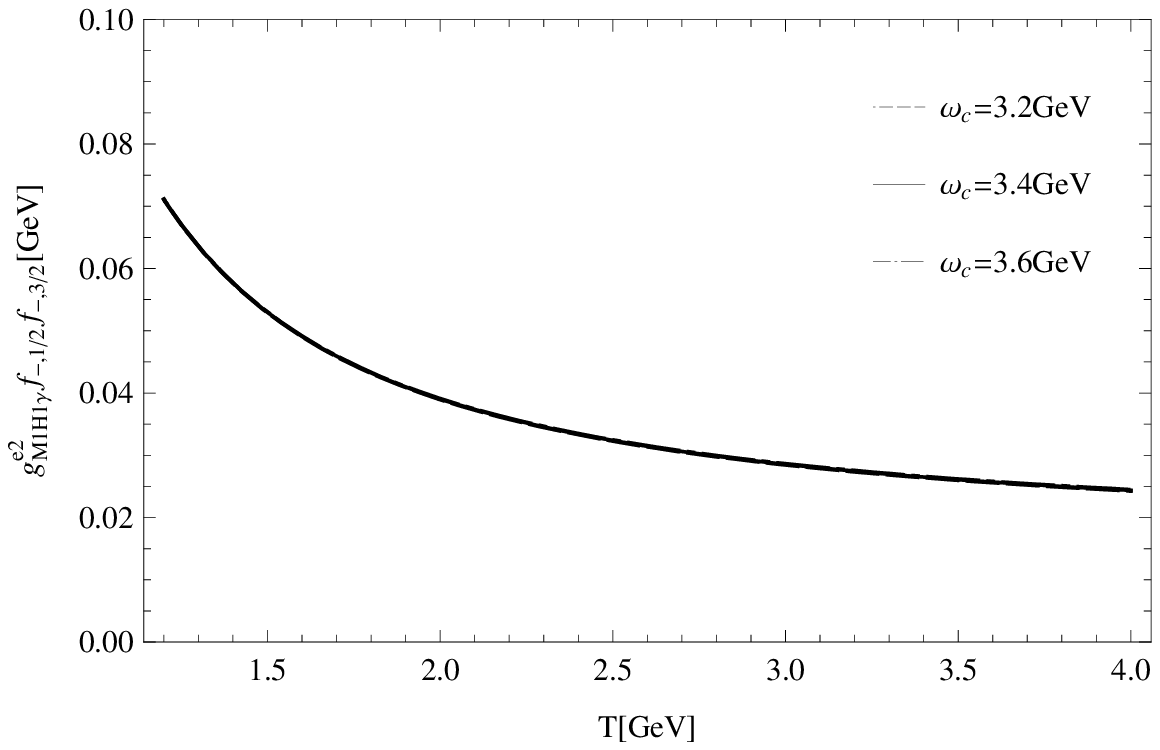}
\setcaptionwidth{3in}
\caption{The sum rule for $g^{e2}_{M_1H_1\gamma}f_{-,1/2}f_{-,3/2}$ with $\omega_c=3.2,3.4,3.6\ \text{GeV}$ and the working interval
$2.0<T<2.4\ \text{GeV}$.} \label{fig:CCM1H1E2gamma}
\end{minipage}
\end{figure}

\begin{figure}
\begin{minipage}[t]{0.5\linewidth}
\centering
\captionstyle{flushleft}
\includegraphics[width=3in]{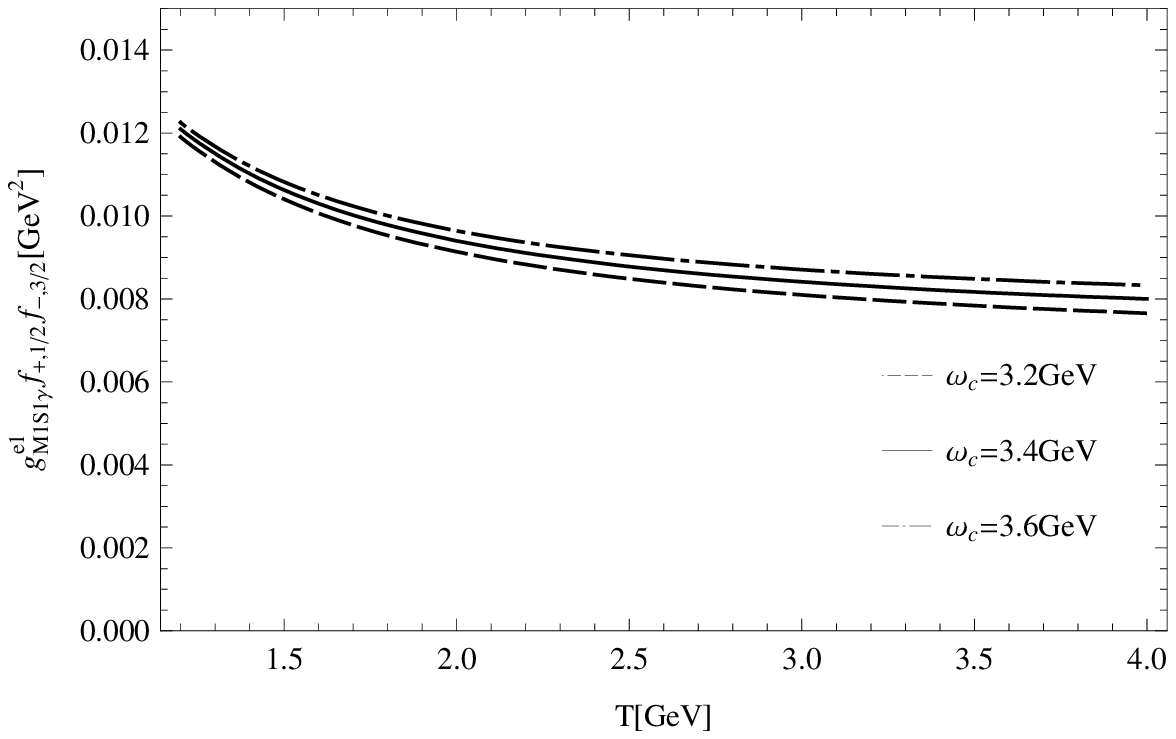}
\setcaptionwidth{3in}
\caption{The sum rule for $g^{e1}_{M_1S_1\gamma}f_{+,1/2}f_{-,3/2}$ with $\omega_c=3.2,3.4,3.6\ \text{GeV}$ and the working interval
$2.5<T<3.5\ \text{GeV}$.} \label{fig:CCM1S1E1gamma}
\end{minipage}%
\begin{minipage}[t]{0.5\linewidth}
\centering
\captionstyle{flushleft}
\includegraphics[width=3in]{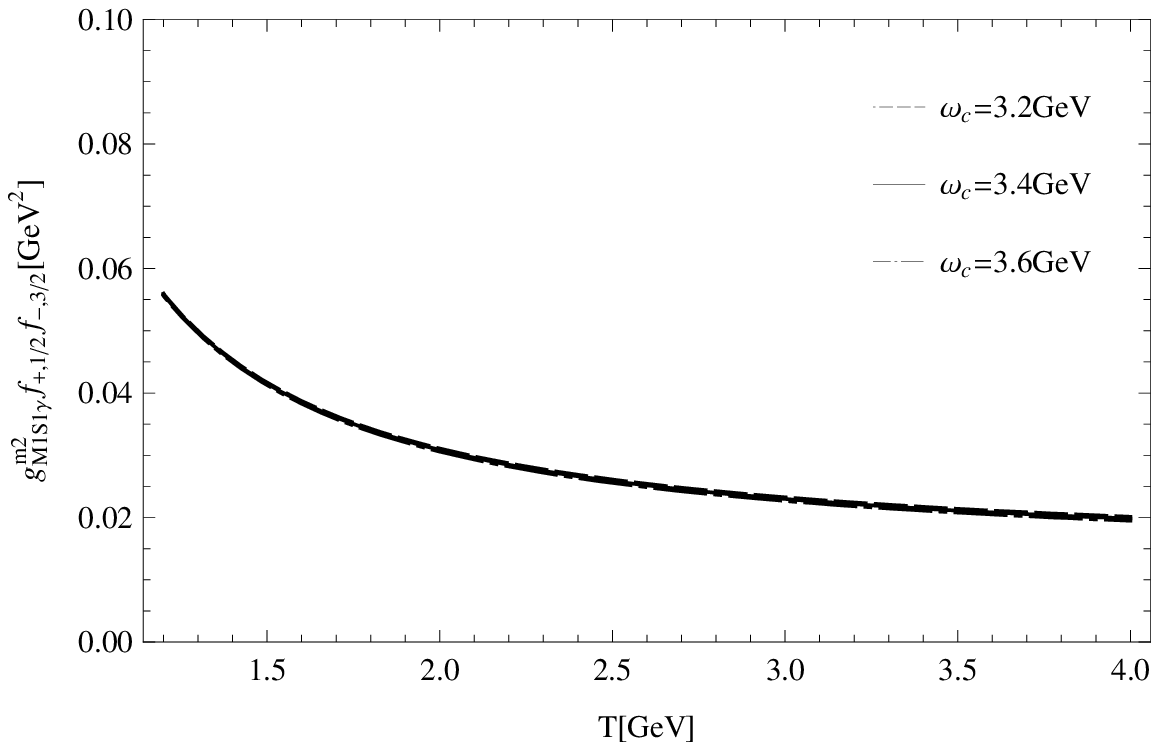}
\setcaptionwidth{3in}
\caption{The sum rule for $g^{m2}_{M_1S_1\gamma}f_{+,1/2}f_{-,3/2}$ with $\omega_c=3.2,3.4,3.6\ \text{GeV}$ and the working interval
$2.5<T<3.5\ \text{GeV}$.} \label{fig:CCM1S1M2gamma}
\end{minipage}
\end{figure}

The resulting sum rules are plotted with $\omega_c=3.2,3.4,3.6\ \text{GeV}$ in Figs. \ref{fig:CCM1H1M1gamma}-\ref{fig:CCM1S1M2gamma}.
%\ref{} should be placed before \begin{figure}?
The numerical values of these coupling constants are collected in Table \ref{tablegammacc}.

The parameters in the distribution amplitudes are
$f_{3 \gamma}=-(0.0039\pm0.0020) \, {\rm GeV}^2$,
$\omega^V_\gamma = 3.8\pm 1.8$, $\omega^A_\gamma = -2.1\pm  1.0$ \cite{gammalcda};
$\kappa=0.2$,  $\zeta_1=0.4$,   $\zeta_2=0.3$,   $\varphi_2=\kappa^+=\zeta_1^+=\zeta_2^+=0$
($\mu=1 $ GeV) \cite{gammaparameter}.
The quark condensate at the same renormalization scale is
$\langle \bar q q \rangle=(-0.245 \,\, {\rm GeV})^3 $($q=u,d$) \cite{rhoparameter}.
We adopt the value $\chi=(3.15 \pm0.3) \,\, {\rm GeV}^{-2} $ \cite{gammalcda} for the magnetic susceptibility  of the quark condensate.

\begin{table}[htb]
\begin{center}
\setlength\extrarowheight{8pt}
\begin{tabular}{ccccccc}
\hline
          & $g^{m1}_{M_1H_1\gamma}$ & $g^{e2}_{M_1H_1\gamma}$ & $g^{e1}_{M_1S_1\gamma}$ & $g^{m2}_{M_1S_1\gamma}$ \\
          & $[\ \text{GeV}^{-1}]$   & $[\ \text{GeV}^{-3}]$   & $[\ \text{GeV}^{-2}]$   & $[\ \text{GeV}^{-2}]$   \\
$g_c$     & $-0.05$                 & $0.37$                  & $0.06$                  & $0.16$                  \\
$g$       & $-0.07\sim-0.03$        & $0.26\sim0.53$          & $0.04\sim0.10$          & $0.10\sim0.28$          \\
\hline
\end{tabular}
\end{center}
\caption{The $\gamma$ coupling constants.
$g_c$s correspond to the central values of the overlap amplitudes and
$\tilde{g}$s with $\omega_c=3.4\ \text{GeV}$ and $T=2.75\ \text{GeV}$($2.2\ \text{GeV}$ for $g^{e2}_{M_1H_1\gamma}$).
The ranges of $g$s are determined according to the uncertainty of the overlap amplitudes
and the ranges of $\tilde{g}$s with $T$ in their working intervals and $3.2<\omega_c<3.6\ \text{GeV}$.
}
\label{tablegammacc}
\end{table}

%\begin{align}
%&g^{m1}_{M_1H_1\gamma}=(-0.04\pm 0.01)\ \text{GeV}^{-1}\,,  &&g^{e2}_{M_1H_1\gamma}=(0.29\pm 0.02)\ \text{GeV}^{-3}\,,\nonumber\\
%&g^{e1}_{M_1S_1\gamma}=(0.05\pm 0.004)\ \text{GeV}^{-2}\,,  &&g^{m2}_{M_1S_1\gamma}=(0.14\pm 0.007)\ \text{GeV}^{-2}\,.
%\end{align}
%\begin{eqnarray}
%&&\tilde{g}^{m1}_{M_1H_1\gamma}=-0.024\pm 0.002\,,\ \tilde{g}^{e2}_{M_1H_1\gamma}=-0.035\pm 0.005\,,\nonumber\\
%&&\tilde{g}^{e1}_{M_1S_1\gamma}=-0.012\pm 0.001\,,\ \tilde{g}^{m2}_{M_1S_1\gamma}=-0.045\pm 0.005\,.
%\end{eqnarray}

%%%%%%%%%%%%%%%%%%%%%%%%%%%%%%%%%%%%%%%%%%%%%%%%
\section{Decay widths}\label{decaywidth}
%%%%%%%%%%%%%%%%%%%%%%%%%%%%%%%%%%%%%%%%%%%%%%%%

It is straightforward to calculate the decay widths of the $(1^-,2^-)$ $D$-wave heavy mesons
with the coupling constants extracted in the previous sections.
For completeness, we make rough estimates of the decay widths of the single and double-pion channels with the given coupling constants,
although some of the coupling constants are extracted from sum sules without a stable working interval,
as shown in Section. \ref{picoupling} and \ref{rhocoupling}.
The formulas for the single-pion and the radiative decay widths are collected in Appendix~\ref{appendixwidths}.

To get the numerical values of these decay widths, we need the mass parameters of the heavy mesons concerned.
They are collected in Table \ref{tablemasses}.
\begin{table}[htb]
\begin{center}
\setlength\extrarowheight{8pt}
\begin{tabular}{c|cccccccc}
\hline
     $J^P$                      &\ $0^-$       &   $1^-$       &   $0^+$       &   $1^+$       &   $1^+$       &   $2^+$       &   $1^-$       &   $2^-$          \\
 $m_D\ \text{[GeV]}$    \ \ \ \ &\ $1.87\ \ $  &   $2.01\ \ $  &   $2.40\ \ $  &   $2.43\ \ $  &   $2.42\ \ $  &   $2.46\ \ $  &   $2.82\ \ $  &   $2.83\ \ $     \\
 $m_B\ \text{[GeV]}$    \ \ \ \ &\ $5.28\ \ $  &   $5.33\ \ $  &   $5.70\ \ $  &   $5.73\ \ $  &   $5.72\ \ $  &   $5.75\ \ $  &   $6.10\ \ $  &   $6.11\ \ $     \\
 $m_{D_s}\ \text{[GeV]}$\ \ \ \ &\ $1.97\ \ $  &   $2.11\ \ $  &&&&&&\\
 $m_{B_s}\ \text{[GeV]}$\ \ \ \ &\ $5.37\ \ $  &   $5.41\ \ $  &&&&&&\\\hline
\end{tabular}
\end{center}
\caption{Heavy meson masses used in our calcultion.
We take the values of $m_D$, $m_{D^*}$, $m_{D^*_0}$, $m_{D_1'}$, $m_{D_1}$, $m_{D_2}$,
$m_B$, $m_{B^*}$, $m_{B_1'}$, $m_{B_1}$, $m_{B_2}$, $m_{D_s}$, and $m_{B_s}$ from Ref. \cite{heavymesonmassesPDG}.
Masses of the other heavy mesons are from the quark model prediction made in Ref. \cite{heavymesonmassesQM}.}
\label{tablemasses}
\end{table}

\begin{table}[htb]
\begin{center}
\setlength\extrarowheight{8pt}
\begin{tabular}{c|cccccccccc|c}
\hline
                           & $D\pi^+$    & $D^*\pi^+$       & $D^*_0\pi^+$       & $D_1'\pi^+$       & $D_1\pi^+$       & $D_2\pi^+$   & $D\eta$   &  $D^*\eta$  & $D_sK^+$  &  $D_s^*K^+$ & $\Gamma_{D\rightarrow D+P}$    \\
$D^{*'}_1\rightarrow$      & 3.7         & 1.3              &                    & 0.02              & 2.1              & 0.04         & 0.3       &  0.1        & 1.6       &  0.4        & 13.1          \\
                           & $1.8-7.9$   & $0.6-2.9$        &                    & $0.01-0.04$       & $0.3-8.9$        & $0.02-0.1$   & $0.2-0.7$ &  $0.1-0.2$  & $0.8-3.4$ &  $0.2-0.8$  & $5.4-34.9$    \\
$D^*_2\rightarrow$         &             & 4.4              & 0.02               & 0.02              & 0.04             & 7.8          &           &  0.3        &           &  1.2        & 19.9          \\
                           &             & $2.1-9.3$        & $0.01-0.05$        & $0.01-0.06$       & $0.02-1.0$       & $1.1-33.5$   &           &  $0.1-0.6$  &           &  $0.6-2.6$  & $5.6-69.1$    \\\hline\hline
                           & $B\pi^+$    & $B^*\pi^+$       & $B^*_0\pi^+$       & $B_1'\pi^+$       & $B_1\pi^+$       & $B_2\pi^+$   & $B\eta$   &  $B^*\eta$  & $B_sK^+$  &  $B_s^*K^+$ &               \\
$B^{*'}_1\rightarrow$      & 4.4         & 1.9              &                    & 0.02              & 2.3              & 0.05         & 0.3       &  0.1        & 1.3       &  0.5        & 15.2          \\
                           & $2.1-9.3$   & $0.9-3.9$        &                    & $0.01-0.04$       & $0.4-9.6$        & $0.02-0.11$  & $0.1-0.6$ &  $0.1-0.2$  & $0.6-2.8$ &  $0.2-1.0$  & $6.1-39.0$    \\
$B^*_2\rightarrow$         &             & 5.8              & 0.02               & 0.02              & 0.05             & 8.5          &           &  0.3        &           &  1.6        & 23.5          \\
                           &             & $2.8-12.4$       & $0.01-0.06$        & $0.01-0.06$       & $0.02-0.11$      & $1.2-36.3$   &           &  $0.2-0.7$  &           &  $0.7-3.3$  & $7.0-77.4$    \\\hline
\end{tabular}
\end{center}
\caption{The widths of the single-pion channels in units of MeV,
including their ranges and the values corresponding to the central
values of $g$s.}
\label{tablewidthspi}         %label should be placed behind the caption to ensure a correct numbering of the tables and/or figures.
\end{table}

The resulting decay widths are presented in Table
\ref{tablewidthspi}. Here we also make rough estimates of the
decay widths of $D+\eta/K^+$ channels assuming the flavor SU(3)
symmetry. The column $\Gamma_{D\rightarrow D+P}$ sums over the
widths of all pseudoscalar meson decay channels, including the
widths of $D\pi^0$ channels which are about half of the widths of
the corresponding $D\pi^+$ channels.

\begin{figure}[hbt]
\begin{center}
\scalebox{0.7}{\includegraphics{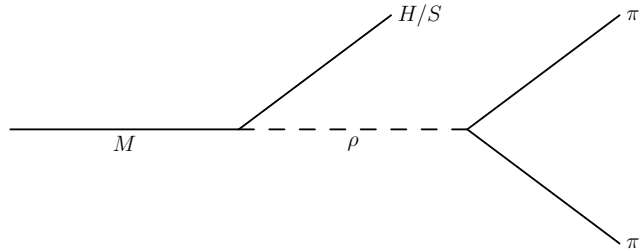}}
 \caption{The double-pion decays of $D$-wave heavy mesons via a virtual $\rho$ meson.} \label{fig:DoublePion}
\end{center}
\end{figure}

The Feynman diagram of the double-pion decays is shown in Fig.
\ref{fig:DoublePion}.
We will not present the expressions for the widths of these
double-pion decay channels due to the length of these expressions.
Their numerical values are collected in Table
\ref{tablewidthsrhogamma}, together with those of the radiative
decay channels. We take $f_{\rho\pi\pi}=6.14$ in our calculation.
\begin{table}[htb]
\begin{center}
\setlength\extrarowheight{8pt}
\begin{tabular}{c|cccc|cccc}
\hline
                        & $D\pi^+\pi^0$ & $D^*\pi^+\pi^0$ & $D^*_0\pi^+\pi^0$ & $D_1'\pi^+\pi^0$ & $D\gamma$  &  $D^*\gamma$  &  $D^*_0\gamma$  &  $D_1'\gamma$   \\
$D^{*'}_1\rightarrow$   & 44.5          & 642.3           & 3.0               & 1.2              & 8.0        &  12.2         &  0.3            & 0.8             \\
                        & $7.1-178.2$   & $241.3-1529.8$  & $1.0-8.5$         & $0.4-3.4$        & $2.9-15.6$ &  $5.6-24.7$   &  $0.1-0.7$      & $0.3-2.3$       \\
$D^*_2\rightarrow$      & 1155.0        & 448.3           & 0.06              & 3.5              & 9.0        &  15.7         &  0.4            & 0.8             \\
                        & $440.6-2728.3$& $165.1-1092.8$  & $0.03-0.1$        & $1.1-10.0$       & $4.4-18.5$ &  $6.5-31.3$   &  $0.2-1.3$      & $0.3-2.3$       \\\hline\hline
                        & $B\pi^+\pi^0$ & $B^*\pi^+\pi^0$ & $B^*_0\pi^+\pi^0$ & $B_1'\pi^+\pi^0$ & $B\gamma$  & $B^*\gamma$   & $B^*_0\gamma$   & $B_1'\gamma$    \\
$B^{*'}_1\rightarrow$   & 19.1          & 579.7           & 2.7               & 1.1              & 9.4        & 18.8          & 0.3             & 0.8             \\
                        & $3.1-76.5$    & $217.3-1382.1$  & $0.8-7.6$         & $0.3-3.0$        & $3.4-18.5$ & $8.8-38.3$    & $0.1-0.8$       & $0.3-2.4$       \\
$B^*_2\rightarrow$      & 521.7         & 414.2           & 0.05              & 3.1              & 9.5        & 22.2          & 0.5             & 0.8             \\
                        & $198.1-1234.1$& $152.5-1008.4$  & $0.02-0.12$       & $1.0-8.7$        & $4.7-19.6$ & $9.3-44.5$    & $0.2-1.4$       & $0.3-2.4$       \\\hline

\end{tabular}
\end{center}
\caption{The widths of the double-pion and radiative decay
channels in units of keV.}
\label{tablewidthsrhogamma}         %label should be placed behind the caption to ensure a correct numbering of the tables and/or figures.
\end{table}

%%%%%%%%%%%%%%%%%%%%%%%%%%%%%%%%%%%%%%%%%%%%%%%%
\section{Conclusion}\label{summary}
%%%%%%%%%%%%%%%%%%%%%%%%%%%%%%%%%%%%%%%%%%%%%%%%

We have calculated the $\pi$, $\rho$, $\omega$, and $\gamma$
couplings with heavy mesons at the leading order of HQET within
the framework of LCQSR. Most of the sum rules obtained are stable
with the variations of the Borel parameter $T$ and the continuum
threshold $\omega_c$. For the other sum rules, we can not find a
stable working interval of $T$. The extracted vector meson heavy
meson coupling constants may be helpful in the study of the
interaction between two $B(D)$ mesons.

Some possible sources of the errors in our calculation include the
inherent inaccuracy of LCQSR: the omission of the higher twist
terms in the OPE near the light-cone, the variation of the
coupling constant with the continuum threshold $\omega_c$ and the
Borel parameter $T$ in the working interval, the omission of the
higher conformal partial waves in the light-cone distribution
amplitudes of the involved mesons or photon, and the uncertainty
in the parameters that appear in these light-cone distribution
amplitudes. The uncertainty in $f$'s and $\bar{\Lambda}$'s is
another source of errors. It's also understood that the $1/m_Q$
correction may turn out to be quite large for the charm mesons
while such a correction is under control for the bottom ssytem.

With these extracted coupling constants, we have made a very rough
estimate of the decay widths of the $(1^-,2^-)$ $D$-wave heavy
mesons. Their dominant decay modes are $M\rightarrow T+\pi$ and
$M\rightarrow H+\pi$. The former mode is of $S$-wave while the
latter decay occurs in the $P$-wave but with a larger phase space.
The total width of the $D$-wave heavy mesons is roughly several
tens of MeV. Therefore the $(1^-,2^-)$ $D$-wave heavy mesons are
not expected to be extremely broad, which is helpful in the future
experimental search.

%%%%%%%%%%%%%%%%%%%%%%%%%%%%%%%%%%%%%%%%
\section*{Acknowledgments}
%%%%%%%%%%%%%%%%%%%%%%%%%%%%%%%%%%%%%%%

This project is supported by the National Natural Science Foundation of China
under Grants No. 10625521, No. 10721063, and the Ministry of Science and
Technology of China (2009CB825200).

%---------------------------------------------------------------------------

\appendix

%%%%%%%%%%%%%%%%%%%%%%%%%%%%%%%%%%%%%%%%%
\section{The definitions of the $\pi$, $\rho$ meson and the photon light-cone distribution amplitudes}\label{appendixLCDA}
%%%%%%%%%%%%%%%%%%%%%%%%%%%%%%%%%%%%%%%%

The 2-particle distribution amplitudes of the $\pi$ meson are defined as \cite{pilcda}
\begin{eqnarray}
\langle 0 | \bar u(z) \gamma_\mu\gamma_5 d(-z) |
  \pi^-(P)\rangle
& = & i f_\pi p_\mu \int_0^1 du\, e^{i\xi pz} \, \phi_\pi(u) +
  \frac{i}{2}\, f_\pi m^2\, \frac{1}{pz}\, z_\mu \int_0^1 du \,
  e^{i\xi pz} g_\pi(u)\,,\label{eq:2.8}\nonumber
\\
\langle 0 | \bar u(z) i\gamma_5 d(-z) | \pi(P)\rangle & = &
\frac{f_\pi m_\pi^2}{m_u+m_d}\, \int_0^1 du \, e^{i\xi pz}\,
\phi_{p}(u)\,,
\label{eq:2.11}\nonumber
\\
\langle 0 | \bar u(z) \sigma_{\alpha\beta}\gamma_5 d(-z) |
\pi(P)\rangle & = &-\frac{i}{3}\, \frac{f_\pi
  m_\pi^2}{m_u+m_d}  (p_\alpha z_\beta-
p_\beta z_\alpha) \int_0^1 du \, e^{i\xi pz}\,\phi_{\sigma}(u)\,,
\label{eq:2.12}
\end{eqnarray}
where $\xi\equiv2u-1$, $\phi_\pi$ is the leading twist-2 distribution amplitude, $\phi_{(p,\sigma)}$ are of twist-3.
All the above distribution amplitudes $\phi=\{\phi_\pi,\phi_p,\phi_\sigma,g_\pi\}$
are normalized to unity: $\int_0^1 du\, \phi(u) = 1$.

There is one 3-particle distribution amplitudes of twist-3, defined as \cite{pilcda}
\begin{eqnarray}
\langle 0 | \bar u(z) \sigma_{\mu\nu}\gamma_5
  gG_{\alpha\beta}(vz) d(-z)| \pi^-(P)\rangle
& = & i\,\frac{f_\pi m_\pi^2}{m_u+m_d} \left(p_\alpha p_\mu
  g_{\nu\beta}^\perp - p_\alpha p_\nu
  g_{\mu\beta}^\perp - p_\beta p_\mu g_{\nu\alpha}^\perp + p_\beta
  p_\nu g_{\alpha\mu}^\perp \right) {\cal T}(v,pz)\,,\label{eq:3pT3}
\end{eqnarray}
where we used the following notation for the integral
defining the 3-particle distribution amplitude:
\begin{equation}
{\cal T}(v,pz) = \int {\cal D}\underline{\alpha} \, e^{-ipz(\alpha_u
  -\alpha_d + v\alpha_g)} {\cal T}(\alpha_d,\alpha_u,\alpha_g)\,.
\end{equation}
Here $\underline{\alpha}$ is the set of three momentum fractions
$\alpha_d$, $\alpha_u$, and $\alpha_g$. The integration measure is
\begin{equation}
\int {\cal D}\underline{\alpha} = \int_0^1 d\alpha_d d\alpha_u
d\alpha_g \delta(1-\alpha_u-\alpha_d-\alpha_g)\,.
\end{equation}
The 3-particle distribution amplitudes of twist-4 are
\begin{eqnarray}
\langle 0 | \bar u(z)\gamma_\mu\gamma_5
gG_{\alpha\beta}(vz)d(-z)|\pi^-(P)\rangle
& = & p_\mu (p_\alpha z_\beta - p_\beta z_\alpha)\, \frac{1}{pz}\, f_\pi
m_\pi^2 {\cal A}_\parallel(v,pz) + (p_\beta g_{\alpha\mu}^\perp -
p_\alpha g_{\beta\mu}^\perp) f_\pi m_\pi^2 {\cal A}_\perp(v,pz)\,,\hspace*{1cm}\nonumber\\
\langle 0 | \bar u(z)\gamma_\mu i
g\widetilde{G}_{\alpha\beta}(vz)d(-z)|\pi^-(P)\rangle\
& = & p_\mu (p_\alpha z_\beta - p_\beta z_\alpha)\, \frac{1}{pz}\, f_\pi
m_\pi^2 {\cal V}_\parallel(v,pz) + (p_\beta g_{\alpha\mu}^\perp -
p_\alpha g_{\beta\mu}^\perp) f_\pi m_\pi^2 {\cal V}_\perp(v,pz)\,,\hspace*{1cm}
\end{eqnarray}
where $\widetilde{G}_{\alpha\beta}$ is the dual field
$\widetilde{G}_{\alpha\beta}\equiv \frac{1}{2}\varepsilon_{\alpha\beta\gamma\delta}G^{\gamma\delta}$.

%%%%%%%%%%%%%%%%%%%%%%%%%%%%%%%%%%%%%%%%

The definitions of the distribution amplitudes of the $\rho$ meson
used in the text read as \cite{rholcda,rhoparameter}
\begin{eqnarray}
\langle 0|\bar u(z) \gamma_{\mu} d(-z)|\rho^-(P,\lambda)\rangle &=&
f_{\rho} m_{\rho} \left[ p_{\mu} \frac{e^{(\lambda)}\cdot z}{p \cdot
z} \int_{0}^{1} \!du\, e^{i \xi p \cdot z} \varphi_{\parallel}(u,
\mu^{2}) \right.
%\nonumber \\
+ e^{(\lambda)}_{\perp \mu}
\int_{0}^{1} \!du\, e^{i \xi p \cdot z} g_{\perp}^{(v)}(u, \mu^{2})
\nonumber \\
& & \left.- \frac{1}{2}z_{\mu}
\frac{e^{(\lambda)}\cdot z }{(p \cdot z)^{2}} m_{\rho}^{2}
\int_{0}^{1} \!du\, e^{i \xi p \cdot z} g_{3}(u, \mu^{2}) \right]\,,\nonumber \\
\langle 0|\bar u(z) \gamma_{\mu} \gamma_{5}
d(-z)|\rho^-(P,\lambda)\rangle &=&
 \frac{1}{2}f_{\rho}
m_{\rho} \varepsilon_{\mu}^{\phantom{\mu}\nu \alpha \beta}
e^{(\lambda)}_{\perp \nu} p_{\alpha} z_{\beta}
\int_{0}^{1} \!du\, e^{i \xi p \cdot z} g^{(a)}_{\perp}(u, \mu^{2})\,,\nonumber
\\
\langle 0|\bar u(z) \sigma_{\mu \nu} d(-z)|\rho^-(P,\lambda)\rangle
&=& i f_{\rho}^{T} \left[ ( e^{(\lambda)}_{\perp \mu}p_\nu -
e^{(\lambda)}_{\perp \nu}p_\mu ) \int_{0}^{1} \!du\, e^{i \xi p
\cdot z} \varphi_{\perp}(u, \mu^{2}) \right.
\nonumber \\
& &+ (p_\mu z_\nu - p_\nu z_\mu )
\frac{e^{(\lambda)} \cdot z}{(p \cdot z)^{2}}
m_{\rho}^{2}
\int_{0}^{1} \!du\, e^{i \xi p \cdot z} h_\parallel^{(t)} (u, \mu^{2})
\nonumber \\
& & \left.+ \frac{1}{2}
(e^{(\lambda)}_{\perp \mu} z_\nu -e^{(\lambda)}_{\perp \nu} z_\mu)
\frac{m_{\rho}^{2}}{p \cdot z}
\int_{0}^{1} \!du\, e^{i \xi p \cdot z} h_{3}(u, \mu^{2}) \right]\,,\nonumber \\
\langle 0|\bar u(z)d(-z)|\rho^-(P,\lambda)\rangle
&=&-if_{\rho}^{T}(e^{(\lambda)}z)m_\rho^2 \int_{0}^{1} \!du\, e^{i \xi p \cdot z} h_\parallel^{(s)} (u, \mu^{2})\,.
\end{eqnarray}
The distribution amplitude $\varphi_\parallel$ and $\varphi_\perp$
are of twist-2, $g_\perp^{(v)}$, $g_\perp^{(a)}$,
$h_\parallel^{(s)}$ and $h_\parallel^{(t)}$ are twist-3 and $g_3$,
$h_3$ are twist-4. All functions
$\phi=\{\varphi_\parallel,\varphi_\perp,
g_\perp^{(v)},g_\perp^{(a)},h_\parallel^{(s)},h_\parallel^{(t)},g_3,h_3\}$
are normalized to satisfy $\int_0^1\!du\, \phi(u) =1$.

The 3-particle distribution amplitudes of the $\rho$ meson are
defined as \cite{rholcda,rhoparameter}
\begin{eqnarray}
\langle 0|\bar u(z) g\widetilde G_{\mu\nu}\gamma_\alpha\gamma_5
  d(-z)|\rho^-(P,\lambda)\rangle &=&
  f_\rho m_\rho p_\alpha[p_\nu e^{(\lambda)}_{\perp\mu}
 -p_\mu e^{(\lambda)}_{\perp\nu}]{\cal A}(v,pz)
\nonumber\\ &&{}
+ f_\rho m_\rho^3\frac{e^{(\lambda)}\cdot z}{pz}
[p_\mu g^\perp_{\alpha\nu}-p_\nu g^\perp_{\alpha\mu}] \widetilde\Phi(v,pz)
\nonumber\\&&{}
+ f_\rho m_\rho^3\frac{e^{(\lambda)}\cdot z}{(pz)^2}
p_\alpha [p_\mu z_\nu - p_\nu z_\mu] \widetilde\Psi(v,pz)\,,\nonumber
\\
\langle 0|\bar u(z) g G_{\mu\nu}i\gamma_\alpha
  d(-z)|\rho^-(P)\rangle &=&
  f_\rho m_\rho p_\alpha[p_\nu e^{(\lambda)}_{\perp\mu}
  - p_\mu e^{(\lambda)}_{\perp\nu}{\cal V}(v,pz)
\nonumber\\&&{}
+ f_\rho m_\rho^3\frac{e^{(\lambda)}\cdot z}{pz}
[p_\mu g^\perp_{\alpha\nu} - p_\nu g^\perp_{\alpha\mu}] \Phi(v,pz)
\nonumber\\&&{}
+ f_\rho m_\rho^3\frac{e^{(\lambda)}\cdot z}{(pz)^2}
p_\alpha [p_\mu z_\nu - p_\nu z_\mu] \Psi(v,pz)\,,\nonumber
\\
\langle 0|\bar u(z) \sigma_{\alpha\beta}
         gG_{\mu\nu}(vz)
         d(-z)|\rho^-(P,\lambda)\rangle
&=& f_{\rho}^T m_{\rho}^3 \frac{e^{(\lambda)}\cdot z }{2 (p \cdot z)}
    [ p_\alpha p_\mu g^\perp_{\beta\nu}
     -p_\beta p_\mu g^\perp_{\alpha\nu}
     -p_\alpha p_\nu g^\perp_{\beta\mu}
     +p_\beta p_\nu g^\perp_{\alpha\mu} ]
      {\cal T}(v,pz)
\nonumber\\
&&+ f_{\rho}^T m_{\rho}^2
    [ p_\alpha e^{(\lambda)}_{\perp\mu}g^\perp_{\beta\nu}
     -p_\beta e^{(\lambda)}_{\perp\mu}g^\perp_{\alpha\nu}
     -p_\alpha e^{(\lambda)}_{\perp\nu}g^\perp_{\beta\mu}
     +p_\beta e^{(\lambda)}_{\perp\nu}g^\perp_{\alpha\mu} ]
      T_1^{(4)}(v,pz)
\nonumber\\
&&+ f_{\rho}^T m_{\rho}^2
    [ p_\mu e^{(\lambda)}_{\perp\alpha}g^\perp_{\beta\nu}
     -p_\mu e^{(\lambda)}_{\perp\beta}g^\perp_{\alpha\nu}
     -p_\nu e^{(\lambda)}_{\perp\alpha}g^\perp_{\beta\mu}
     +p_\nu e^{(\lambda)}_{\perp\beta}g^\perp_{\alpha\mu} ]
      T_2^{(4)}(v,pz)
\nonumber\\
&&+ \frac{f_{\rho}^T m_{\rho}^2}{pz}
    [ p_\alpha p_\mu e^{(\lambda)}_{\perp\beta}z_\nu
     -p_\beta p_\mu e^{(\lambda)}_{\perp\alpha}z_\nu
     -p_\alpha p_\nu e^{(\lambda)}_{\perp\beta}z_\mu
     +p_\beta p_\nu e^{(\lambda)}_{\perp\alpha}z_\mu ]
      T_3^{(4)}(v,pz)
\nonumber\\
&&+ \frac{f_{\rho}^T m_{\rho}^2}{pz}
    [ p_\alpha p_\mu e^{(\lambda)}_{\perp\nu}z_\beta
     -p_\beta p_\mu e^{(\lambda)}_{\perp\nu}z_\alpha
     -p_\alpha p_\nu e^{(\lambda)}_{\perp\mu}z_\beta
     +p_\beta p_\nu e^{(\lambda)}_{\perp\mu}z_\alpha]
      T_4^{(4)}(v,pz)\,,\nonumber
\\
\langle 0|\bar u(z)
         gG_{\mu\nu}(vz)
         d(-z)|\rho^-(P,\lambda)\rangle
&=& i f_{\rho}^T m_{\rho}^2
 [e^{(\lambda)}_{\perp\mu}p_\nu-e^{(\lambda)}_{\perp\nu}p_\mu] S(v,pz)\,,
\nonumber\\
\langle 0|\bar u(z)
         ig\widetilde G_{\mu\nu}(vz)\gamma_5
         d(-z)|\rho^-(P,\lambda)\rangle
&=& i f_{\rho}^T m_{\rho}^2
 [e^{(\lambda)}_{\perp\mu}p_\nu-e^{(\lambda)}_{\perp\nu}p_\mu]
  \widetilde S(v,pz)\,.
\end{eqnarray}
The distribution amplitudes ${\cal A}$, ${\cal V}$ and ${\cal T}$ are of
twist-3 and the other 3-particle distribution amplitudes are of twist-4.

%%%%%%%%%%%%%%%%%%%%%%%%%%%%%%%%%%%%%%%%

The definitions of the 2-particle distribution amplitudes of the
photon read as\cite{gammalcda}
\begin{eqnarray}
\langle 0|\bar q(z) \sigma_{\alpha\beta} q(-z)
   |\gamma^{(\lambda)}(q)\rangle &=&
 i  \,e_q\,e\,\chi\, \langle\bar q q\rangle
 \left( q_\beta e^{(\lambda)}_\alpha- q_\alpha e^{(\lambda)}_\beta \right)
 \int\limits_0^1 \!du\, e^{i\xi qz}\, \phi_{\gamma}(u,\mu)
\nonumber\\
&&+\frac i2  e_q\,e\,\frac{\langle\bar q q\rangle}{qz}
 \left( z_\beta e^{(\lambda)}_\alpha -z_\alpha e^{(\lambda)}_\beta \right)
 \int\limits_0^1 \!du\, e^{i\xi qz}\, h_{\gamma}(u,\mu)\,,\nonumber\\
 \langle 0 |\bar q (z) \gamma_{\mu} q (-z)| \gamma^{(\lambda)}(q)\rangle
 &= & e_q\,e\, f_{3\gamma}\, e^{(\lambda)}_{\perp\mu}
\int_{0}^{1} \!du\,e^{i\xi qz}\, \psi^{(v)}(u, \mu)\,.
\nonumber\\
\langle 0|\bar q(z) \gamma_{\mu} \gamma_{5} q(-z)| \gamma^{(\lambda)}(q)
 \rangle
&=&-\frac12e_q\,e\, f_{3\gamma}\, \varepsilon_{\mu \nu q z}\,e^{(\lambda)}_{\perp\nu}
\int_{0}^{1} \!du\, e^{i\xi qz}\,\psi^{(a)}(u, \mu)\,,
\end{eqnarray}
The 3-particle distribution amplitudes of the photon are defined
as \cite{gammalcda}
\begin{eqnarray}
\langle 0
|\bar q(z) g\widetilde G_{\mu\nu}(vz)\gamma_\alpha\gamma_5
  q(-z)|\gamma^{(\lambda)}(q) \rangle  &=& e_q \,e\,f_{3\gamma}\,
q_\alpha [q_\nu e^{(\lambda)}_{\perp\mu}
  - q_\mu e^{(\lambda)}_{\perp\nu}]
{\cal A}(v,qz)\,,\nonumber
\\
\langle 0 |\bar q(z) g G_{\mu\nu}(vz)i\gamma_\alpha
  q(-z)|\gamma^{(\lambda)}(q) \rangle  &=& e_q \,e\,f_{3\gamma}\,
q_\alpha[q_\nu e^{(\lambda)}_{\perp\mu}
  - q_\mu e^{(\lambda)}_{\perp\nu}]
{\cal V}(v,qz)\,,\nonumber
\\
\langle 0 | \bar q(z)g{G}_{\mu\nu}(vz) q(-z) | \gamma^{(\lambda)}(q) \rangle
&=&
ie_q \,e\,\langle\bar q q\rangle
[q_\nu e^{(\lambda)}_{\perp\mu}-q_\mu e^{(\lambda)}_{\perp\nu}]
S(v,qz)\,,
\nonumber\\
\langle 0|
\bar q(z)g\tilde{G}_{\mu\nu}(vz)i\gamma_5 q(-z) |\gamma^{(\lambda)}(q) \rangle
&=&
ie_{q}\,e\,\langle\bar q q\rangle [q_\nu e^{(\lambda)}_{\perp\mu} -q_\mu e^{(\lambda)}_{\perp\nu}]
\widetilde{S}(v,qz)\,,\nonumber\\
\langle 0 |
\bar q(z)\sigma_{\alpha \beta}g{G}_{\mu\nu}(vz) q(-z) |\gamma^{(\lambda)}(q) \rangle&=&
e_{q}\,e\,\langle\bar q q\rangle
[ q_\alpha e^{(\lambda)}_{\perp\mu}g^\perp_{\beta\nu}
     -q_\beta e^{(\lambda)}_{\perp\mu}g^\perp_{\alpha\nu}
     -q_\alpha e^{(\lambda)}_{\perp\nu}g^\perp_{\beta\mu}
     +q_\beta e^{(\lambda)}_{\perp\nu}g^\perp_{\alpha\mu} ]
T_1(v,qz)
\nonumber\\
&&+
e_{q}\,e\,\langle\bar q q\rangle
 [ q_\mu e^{(\lambda)}_{\perp\alpha}g^\perp_{\beta\nu}
     -q_\mu e^{(\lambda)}_{\perp\beta}g^\perp_{\alpha\nu}
     -q_\nu e^{(\lambda)}_{\perp\alpha}g^\perp_{\beta\mu}
     +q_\nu e^{(\lambda)}_{\perp\beta}g^\perp_{\alpha\mu} ]
      T_2(v,qz)
\nonumber \\
&&+ \frac{e_q\,e\,\langle\bar q q\rangle}{qz}
[ q_\alpha q_\mu e^{(\lambda)}_{\perp\beta}z_\nu
     -q_\beta q_\mu e^{(\lambda)}_{\perp\alpha}z_\nu
     -q_\alpha q_\nu e^{(\lambda)}_{\perp\beta}z_\mu
     +q_\beta q_\nu e^{(\lambda)}_{\perp\alpha}z_\mu ]
      T_3(v,qz)
\nonumber\\
&&+ \frac{e_q\,e\,\langle\bar q q\rangle}{qz}
    [ q_\alpha q_\mu e^{(\lambda)}_{\perp\nu}z_\beta
     -q_\beta q_\mu e^{(\lambda)}_{\perp\nu}z_\alpha
     -q_\alpha q_\nu e^{(\lambda)}_{\perp\mu}z_\beta
     +q_\beta q_\nu e^{(\lambda)}_{\perp\mu}z_\alpha ]
      T_4(v,qz)\,.\hspace*{1cm}\nonumber\\
\end{eqnarray}
Here $\phi_{\gamma}$ is the leading twist-2 distribution amplitude, $\psi^{(v)}$, $\psi^{(a)}$, ${\cal A}$, and ${\cal V}$ are of twist-3,
$h_{\gamma}$, $S$, $\widetilde{S}$, and $T_{(1,2,3,4)}$ are of twist-4.

\section{The definitions of $\mathcal {F}^{[n]}$ and $\mathcal {F}^{[m,n]}$}\label{appendixF}
%%%%%%%%%%%%%%%%%%%%%%%%%%%%%%%%%%%%%%%%

The functions $\mathcal {F}^{[n]}$ and $\mathcal {F}^{[m,n]}$ used in the text are defined as
\begin{eqnarray}
\mathcal {F}^{[n]}(\bar{u}_0)&\equiv&\int_0^{\bar{u}_0}\cdots \int_0^{x_3}\int_0^{x_2}\mathcal {F}(x_1)dx_1dx_2\cdots dx_n\,,\nonumber
\\
\mathcal {F}^{[0,0]}(u_0)&\equiv&\int_0^{u_0}\int_{u_0-\alpha_2}^{1-\alpha_2}\frac{\mathcal {F}(1-\alpha_2-\alpha_3,\alpha_2,\alpha_3)}{\alpha_3}d\alpha_3d\alpha_2\,,\nonumber
\\
\mathcal {F}^{[0,1]}(u_0)&\equiv&\int_0^{u_0}\int_{u_0-\alpha_2}^{1-\alpha_2}\frac{(u_0-\alpha_2)\mathcal {F}(1-\alpha_2-\alpha_3,\alpha_2,\alpha_3)}{\alpha_3^2}d\alpha_3d\alpha_2\,,\nonumber
\\
\mathcal {F}^{[1,0]}(u_0)&\equiv&\int_0^{u_0}\frac{\mathcal {F}(1-u_0,\alpha_2,u_0-\alpha_2)}{u_0-\alpha_2}d\alpha_2
-\int_0^{1-u_0}\frac{\mathcal {F}(1-u_0-\alpha_3,u_0,\alpha_3)}{\alpha_3}d\alpha_3\,,\nonumber
\\
\mathcal {F}^{[1,1]}(u_0)
&\equiv&\int_0^{u_0}\frac{\mathcal {F}(1-u_0,\alpha_2,u_0-\alpha_2)}{u_0-\alpha_2}d\alpha_2
-\int_0^{u_0}\int_{u_0-\alpha_2}^{1-\alpha_2}\frac{\mathcal {F}(1-\alpha_2-\alpha_3,\alpha_2,\alpha_3)}{\alpha_3^2}d\alpha_3d\alpha_2\,,\nonumber
\\
\mathcal {F}^{[2,0]}(u_0)
&\equiv&\frac{\mathcal {F}(0,u_0,1-u_0)}{1-u_0}+\frac{\mathcal {F}(1-u_0-\alpha_3,u_0,\alpha_3)}{\alpha_3}\biggl\lvert_{\alpha_3=0}\nonumber\\
&&+\int_0^{u_0}d\alpha_2\frac{\partial[\mathcal {F}(1-\alpha_2-\alpha_3,\alpha_2,\alpha_3)/\alpha_3]}{\partial\alpha_3}\biggl\lvert_{\alpha_3=u_0-\alpha_2}\nonumber\\
&&-\int_0^{1-u_0}d\alpha_3\frac{\partial[\mathcal {F}(1-\alpha_2-\alpha_3,\alpha_2,\alpha_3)]/\partial\alpha_2}{\alpha_3}\biggl\lvert_{\alpha_2=u_0}\,,\nonumber
\\
\mathcal {F}^{[2,1]}(u_0)
&\equiv&\frac{\mathcal {F}(1-u_0-\alpha_3,u_0,\alpha_3)}{\alpha_3}\biggl\lvert_{\alpha_3=0}
+\int_0^{u_0}d\alpha_2\frac{\partial[\mathcal {F}(1-\alpha_2-\alpha_3,\alpha_2,\alpha_3)/\alpha_3]}{\partial\alpha_3}\biggl\lvert_{\alpha_3=u_0-\alpha_2}\nonumber\\
&&+\int_0^{u_0}d\alpha_2\frac{\mathcal {F}(1-u_0,\alpha_2,u_0-\alpha_2)}{(u_0-\alpha_2)^2}
-\int_0^{1-u_0}d\alpha_3\frac{\mathcal {F}(1-u_0-\alpha_3,u_0,\alpha_3)}{\alpha_3^2}\,,\nonumber
\\
\mathcal {F}^{[3,0]}(u_0)
&\equiv&\frac{\partial[\mathcal {F}(1-\alpha_2-\alpha_3,\alpha_2,\alpha_3)/\partial\alpha_2]}{\alpha_3}\biggl\lvert_{\alpha_2=u_0,\alpha_3=0}
+\frac{\partial[\mathcal {F}(1-\alpha_2-\alpha_3,\alpha_2,\alpha_3)/\alpha_3]}{\partial\alpha_3}\biggl\lvert_{\alpha_2=u_0,\alpha_3=0}\nonumber\\
&&+\frac{\partial[\mathcal {F}(0,1-\alpha_3,\alpha_3)/\alpha_3]}{\partial\alpha_3}\biggl\lvert_{\alpha_3=1-u_0}
+\frac{\partial[\mathcal {F}(u_0-\alpha_2,\alpha_2,1-u_0)/(1-u_0)]}{\partial\alpha_2}\biggl\lvert_{\alpha_2=u_0}\nonumber\\
&&+\int_0^{u_0}d\alpha_2\frac{\partial^2[\mathcal {F}(1-\alpha_2-\alpha_3,\alpha_2,\alpha_3)/\alpha_3]}{\partial\alpha_3^2}\biggl\lvert_{\alpha_3=u_0-\alpha_2}\nonumber\\
&&-\int_0^{1-u_0}d\alpha_3\frac{\partial^2[\mathcal {F}(1-\alpha_2-\alpha_3,\alpha_2,\alpha_3)]/\partial\alpha_2^2}{\alpha_3}\biggl\lvert_{\alpha_2=u_0}\,,\nonumber
\\
\mathcal {F}^{[3,1]}(u_0)
&\equiv&\frac{\partial[\mathcal {F}(1-\alpha_2-\alpha_3,\alpha_2,\alpha_3)/\partial\alpha_2]}{\alpha_3}\biggl\lvert_{\alpha_2=u_0,\alpha_3=0}
+\frac{\partial[\mathcal {F}(1-\alpha_2-\alpha_3,\alpha_2,\alpha_3)/\alpha_3]}{\partial\alpha_3}\biggl\lvert_{\alpha_2=u_0,\alpha_3=0}\nonumber\\
&&+\frac{\mathcal {F}(1-\alpha_2-\alpha_3,\alpha_2,\alpha_3)}{\alpha_3^2}\biggl\lvert_{\alpha_2=u_0,\alpha_3=0}
+\frac{\mathcal {F}(1-\alpha_2-\alpha_3,\alpha_2,\alpha_3)}{\alpha_3^2}\biggl\lvert_{\alpha_2=u_0,\alpha_3=1-u_0}\nonumber\\
&&+\int_0^{u_0}d\alpha_2\frac{\partial^2[\mathcal {F}(1-\alpha_2-\alpha_3,\alpha_2,\alpha_3)/\alpha_3]}{\partial\alpha_3^2}\biggl\lvert_{\alpha_3=u_0-\alpha_2}\nonumber\\
&&+\int_0^{u_0}d\alpha_2\frac{\partial[\mathcal {F}(1-\alpha_2-\alpha_3,\alpha_2,\alpha_3)/\alpha_3^2]}{\partial\alpha_3}\biggl\lvert_{\alpha_3=u_0-\alpha_2}\nonumber\\
&&-\int_0^{1-u_0}d\alpha_3\frac{\partial\mathcal {F}(1-\alpha_2-\alpha_3,\alpha_2,\alpha_3)/\partial\alpha_2}{\alpha_3^2}\biggl\lvert_{\alpha_2=u_0}\,,\nonumber
\\
\mathcal {F}^{[-1,0]}(u_0)
&\equiv&\int_0^{u_0}\int_0^{u_0-\alpha_2}\mathcal {F}(1-\alpha_2-\alpha_3,\alpha_2,\alpha_3)d\alpha_3d\alpha_2\nonumber\\
&&+\int_0^{u_0}\int_{u_0-\alpha_2}^{1-\alpha_2}\frac{(u_0-\alpha_2)\mathcal {F}(1-\alpha_2-\alpha_3,\alpha_2,\alpha_3)}{\alpha_3}d\alpha_3d\alpha_2\,,\nonumber
\\
\mathcal {F}^{[-1,1]}(u_0)
&\equiv&\frac{1}{2}\int_0^{u_0}\int_0^{u_0-\alpha_2}\mathcal {F}(1-\alpha_2-\alpha_3,\alpha_2,\alpha_3)d\alpha_3d\alpha_2\nonumber\\
&&+\frac{1}{2}\int_0^{u_0}\int_{u_0-\alpha_2}^{1-\alpha_2}\frac{(u_0-\alpha_2)^2\mathcal {F}(1-\alpha_2-\alpha_3,\alpha_2,\alpha_3)}{\alpha_3^2}d\alpha_3d\alpha_2\,,\nonumber
\\
\mathcal {F}^{[-2,0]}(u_0)
&\equiv&\int_0^{u_0}\int_0^{u_0-\alpha_2}\int_0^{\alpha_3}\mathcal {F}(1-\alpha_2-x,\alpha_2,x)dxd\alpha_3d\alpha_2\nonumber\\
&&+\frac{1}{2}\int_0^{u_0}\int_0^{u_0-\alpha_2}\alpha_3\mathcal {F}(1-\alpha_2-\alpha_3,\alpha_2,\alpha_3)d\alpha_3d\alpha_2\nonumber\\
&&+\frac{1}{2}\int_0^{u_0}\int_{u_0-\alpha_2}^{1-\alpha_2}\frac{(u_0-\alpha_2)^2}{\alpha_3}\mathcal {F}(1-\alpha_2-\alpha_3,\alpha_2,\alpha_3)d\alpha_3d\alpha_2\,,\nonumber
\\
\mathcal {F}^{[-3,0]}(u_0)
&\equiv&\int_0^{u_0}\int_0^{u_0-\alpha_2}\int_0^{\alpha_3}\int_0^{x_2}\mathcal {F}(1-\alpha_2-x_1,\alpha_2,x_1)dx_1dx_2d\alpha_3d\alpha_2\nonumber\\
&&+\frac{1}{2}\int_0^{u_0}\int_0^{u_0-\alpha_2}\int_0^{\alpha_3}x\mathcal {F}(1-\alpha_2-x,\alpha_2,x)dxd\alpha_3d\alpha_2\nonumber\\
&&+\frac{1}{6}\int_0^{u_0}\int_0^{u_0-\alpha_2}\alpha_3^2\mathcal {F}(1-\alpha_2-\alpha_3,\alpha_2,\alpha_3)d\alpha_3d\alpha_2\nonumber\\
&&+\frac{1}{6}\int_0^{u_0}\int_{u_0-\alpha_2}^{1-\alpha_2}\frac{(u_0-\alpha_2)^3}{\alpha_3}\mathcal {F}(1-\alpha_2-\alpha_3,\alpha_2,\alpha_3)d\alpha_3d\alpha_2\,.
\end{eqnarray}

%%%%%%%%%%%%%%%%%%%%%%%%%%%%%%%%%%%%%%%%%
\section{The formulas for the single-pion and the radiative decay widths}\label{appendixwidths}
%%%%%%%%%%%%%%%%%%%%%%%%%%%%%%%%%%%%%%%%%

The single-pion decay widths can be expressed in terms of the 3-momentum $\bm{q}$ of the final pion:
\begin{eqnarray}
\Gamma(M_1 \rightarrow H_0 \pi^+)&=&\frac{1}{24\pi} \frac{m_{H_0}}{m_{M_1}} (g^{p1}_{M_1H_0\pi})^2 |\bm{q}|^3\,,\nonumber\\
\Gamma(M_1 \rightarrow H_1 \pi^+)&=&\frac{1}{12\pi} \frac{m_{H_1}}{m_{M_1}} (g^{p1}_{M_1H_1\pi})^2 |\bm{q}|^3\,,\nonumber\\
\Gamma(M_1 \rightarrow S_1 \pi^+)&=&\frac{1}{36\pi} \frac{m_{S_1}}{m_{M_1}} (g^{d2}_{M_1S_1\pi})^2 |\bm{q}|^5
                                   +\frac{1}{54\pi} \frac{1}{m_{M_1}m_{S_1}}(g^{d2}_{M_1S_1\pi})^2 |\bm{q}|^7\,,\nonumber\\
\Gamma(M_1 \rightarrow T_1 \pi^+)&=&\frac{1}{8\pi}  \frac{m_{T_1}}{m_{M_1}} (g^{s0}_{M_1T_1\pi})^2 |\bm{q}|
                                   +\frac{1}{24\pi} \frac{1}{m_{M_1}m_{T_1}}(g^{s0}_{M_1T_1\pi})^2 |\bm{q}|^3
                                   -\frac{1}{18\pi} \frac{g^{s0}_{M_1T_1\pi}g^{d2}_{M_1T_1\pi}}{m_{M_1}m_{T_1}} |\bm{q}|^5\nonumber\\
                                 &&+\frac{1}{36\pi} \frac{m_{T_1}}{m_{M_1}} (g^{d2}_{M_1T_1\pi})^2 |\bm{q}|^5
                                   +\frac{1}{54\pi} \frac{1}{m_{M_1}m_{T_1}}(g^{d2}_{M_1T_1\pi})^2 |\bm{q}|^7\,,\nonumber\\
\Gamma(M_1 \rightarrow T_2 \pi^+)&=&\frac{1}{6\pi}  \frac{m_{T_2}}{m_{M_1}} (g^{d2}_{M_1T_1\pi})^2 |\bm{q}|^5
                                   +\frac{1}{6\pi}  \frac{1}{m_{M_1}m_{T_2}}(g^{d2}_{M_1T_1\pi})^2 |\bm{q}|^7\,,\nonumber\\
\Gamma(M_2 \rightarrow H_1 \pi^+)&=&\frac{1}{24\pi} \frac{m_{H_1}}{m_{M_2}} (g^{p1}_{M_2H_1\pi})^2 |\bm{q}|^3
                                   +\frac{1}{60\pi} \frac{1}{m_{M_2}m_{H_1}}(g^{p1}_{M_2H_1\pi})^2 |\bm{q}|^5\,,\nonumber\\
\Gamma(M_2 \rightarrow S_0 \pi^+)&=&\frac{1}{60\pi} \frac{m_{S_0}}{m_{M_2}} (g^{d2}_{M_2S_0\pi})^2 |\bm{q}|^5\,,\nonumber\\
\Gamma(M_2 \rightarrow S_1 \pi^+)&=&\frac{1}{40\pi} \frac{m_{S_1}}{m_{M_2}} (g^{d2}_{M_2S_1\pi})^2 |\bm{q}|^5\,,\nonumber\\
\Gamma(M_2 \rightarrow T_1 \pi^+)&=&\frac{1}{10\pi} \frac{m_{T_1}}{m_{M_2}} (g^{d2}_{M_2T_1\pi})^2 |\bm{q}|^5\,,\nonumber\\
\Gamma(M_2 \rightarrow T_2 \pi^+)&=&\frac{1}{2\pi}    \frac{m_{T_2}}{m_{M_2}}    (g^{s0}_{M_2T_2\pi})^2 |\bm{q}|
                                   +\frac{1}{3\pi}    \frac{1}{m_{M_2}m_{T_2}}   (g^{s0}_{M_2T_2\pi})^2 |\bm{q}|^3
                                   -\frac{2}{45\pi}   \frac{1}{m_{M_2}m_{T_2}^3} (g^{s0}_{M_2T_2\pi})^2|\bm{q}|^5\nonumber\\
                                 &&+\frac{7}{30\pi}   \frac{g^{s0}_{M_2T_2\pi}g^{d2}_{M_2T_2\pi}}{m_{M_2}m_{T_2}}    |\bm{q}|^5
                                   +\frac{7}{80\pi}   \frac{m_{T_2}}{m_{M_2}}    (g^{d2}_{M_2T_2\pi})^2 |\bm{q}|^5
                                   +\frac{2}{45\pi}   \frac{g^{s0}_{M_2T_2\pi}g^{d2}_{M_2T_2\pi}}{m_{M_2}m_{T_2}^3}  |\bm{q}|^7\nonumber\\
                                 &&+\frac{11}{240\pi} \frac{1}{m_{M_2}m_{T_2}}   (g^{d2}_{M_2T_2\pi})^2 |\bm{q}|^7
                                   +\frac{1}{90\pi}   \frac{1}{m_{M_2}m_{T_2}^3} (g^{d2}_{M_2T_2\pi})^2 |\bm{q}|^9\,.
\end{eqnarray}

The above formulas for the decays $M_{(1,2)} \rightarrow H_{(0,1)}+\pi^+$ are also valid for
the decays $M_{(1,2)} \rightarrow H_{(0,1)}+K^+/\eta$, except for an extra $1/6$ isospin factor in the case of $\eta$ decays.

Similarly, the radiative decay widths read as
\begin{eqnarray}
\Gamma(M_1 \rightarrow H_0 \gamma)&=&\frac{e^2}{12\pi} \frac{m_{H_0}}{m_{M_1}}  (g^{m1}_{M_1H_0\gamma})^2 |\bm{q}|^3\,,\nonumber\\
\Gamma(M_1 \rightarrow H_1 \gamma)&=&\frac{e^2}{6\pi}  \frac{m_{H_1}}{m_{M_1}}  (g^{m1}_{M_1H_1\gamma})^2|\bm{q}|^3
                                    +\frac{e^2}{12\pi} \frac{1}{m_{M_1}m_{H_1}} (g^{m1}_{M_1H_1\gamma})^2|\bm{q}|^5
                                    +\frac{e^2}{12\pi} \frac{g^{m1}_{M_1H_1\gamma}g^{e2}_{M_1H_1\gamma}}{m_{M_1}m_{H_1}} |\bm{q}|^7\nonumber\\
                                  &&+\frac{e^2}{24\pi} \frac{m_{H_1}}{m_{M_1}}  (g^{e2}_{M_1H_1\gamma})^2|\bm{q}|^7
                                    +\frac{e^2}{48\pi} \frac{1}{m_{M_1}m_{H_1}} (g^{e2}_{M_1H_1\gamma})^2|\bm{q}|^9\,,\nonumber\\
\Gamma(M_1 \rightarrow S_0 \gamma)&=&\frac{e^2}{12\pi} \frac{m_{S_0}}{m_{M_1}}  (g^{e1}_{M_1S_0\gamma})^2 |\bm{q}|^5\,,\nonumber\\
\Gamma(M_1 \rightarrow S_1 \gamma)&=&\frac{e^2}{6\pi}  \frac{m_{S_1}}{m_{M_1}}  (g^{e1}_{M_1S_1\gamma})^2|\bm{q}|^5
                                    +\frac{e^2}{6\pi}  \frac{m_{S_1}}{m_{M_1}}  (g^{m2}_{M_1S_1\gamma})^2|\bm{q}|^5
                                    +\frac{e^2}{12\pi} \frac{1}{m_{M_1}m_{S_1}} (g^{e1}_{M_1S_1\gamma})^2|\bm{q}|^7\nonumber\\
                                  &&-\frac{e^2}{6\pi} \frac{g^{m2}_{M_1S_1\gamma}g^{e1}_{M_1S_1\gamma}}{m_{M_1}m_{S_1}} |\bm{q}|^7
                                    +\frac{e^2}{12\pi} \frac{1}{m_{M_1}m_{S_1}} (g^{m2}_{M_1S_1\gamma})^2|\bm{q}|^7\,,\nonumber\\
\Gamma(M_2 \rightarrow H_0 \gamma)&=&\frac{e^2}{40\pi} \frac{m_{H_0}}{m_{M_2}}  (g^{e2}_{M_2H_0\gamma})^2 |\bm{q}|^7\;,\nonumber\\
\Gamma(M_2 \rightarrow H_1 \gamma)&=&\frac{e^2}{3\pi}  \frac{m_{H_1}}{m_{M_2}}  (g^{m1}_{M_2H_1\gamma})^2|\bm{q}|^3
                                    +\frac{e^2}{10\pi} \frac{1}{m_{M_2}m_{H_1}} (g^{m1}_{M_2H_1\gamma})^2|\bm{q}|^5
                                    +\frac{e^2}{10\pi} \frac{g^{m1}_{M_2H_1\gamma}g^{e2}_{M_2H_1\gamma}}{m_{M_2}m_{H_1}} |\bm{q}|^7\nonumber\\
                                  &&+\frac{3e^2}{20\pi}\frac{m_{H_1}}{m_{M_2}}  (g^{e2}_{M_2H_1\gamma})^2|\bm{q}|^7
                                    +\frac{e^2}{40\pi} \frac{1}{m_{M_2}m_{H_1}} (g^{e2}_{M_2H_1\gamma})^2|\bm{q}|^9\,,\nonumber\\
\Gamma(M_2 \rightarrow S_0 \gamma)&=&\frac{e^2}{10\pi} \frac{m_{S_0}}{m_{M_2}}  (g^{m2}_{M_2S_0\gamma})^2 |\bm{q}|^5\,,\nonumber\\
\Gamma(M_2 \rightarrow S_1 \gamma)&=&\frac{e^2}{3\pi}  \frac{m_{S_1}}{m_{M_2}}  (g^{e1}_{M_2S_1\gamma})^2|\bm{q}|^5
                                    +\frac{3e^2}{5\pi} \frac{m_{S_1}}{m_{M_2}}  (g^{m2}_{M_2S_1\gamma})^2|\bm{q}|^5
                                    +\frac{e^2}{10\pi} \frac{1}{m_{M_2}m_{S_1}} (g^{e1}_{M_1S_1\gamma})^2|\bm{q}|^7\nonumber\\
                                  &&-\frac{e^2}{5\pi} \frac{g^{m2}_{M_2S_1\gamma}g^{e1}_{M_2S_1\gamma}}{m_{M_2}m_{S_1}} |\bm{q}|^7
                                    +\frac{e^2}{10\pi} \frac{1}{m_{M_2}m_{S_1}} (g^{m2}_{M_2S_1\gamma})^2|\bm{q}|^7\,.
\end{eqnarray}

\end{document}